\documentclass[fleqn,usenatbib]{mnras}

\usepackage{newtxtext,newtxmath}

\usepackage[T1]{fontenc}
\usepackage{multicol}
\usepackage{longtable}

\DeclareRobustCommand{\VAN}[3]{#2}
\let\VANthebibliography\thebibliography
\def\thebibliography{\DeclareRobustCommand{\VAN}[3]{##3}\VANthebibliography}

\usepackage{graphicx}	
\usepackage{amsmath}	

\title[The Type II GC NGC\,1851]{A deep dive into the Type II Globular Cluster NGC\,1851}

\author[E. Dondoglio et al.]{E. Dondoglio,$^{1}$\thanks{E-mail: emanuele.dondoglio@phd.unipd.it}
A. P. Milone,$^{1,2}$
A. F. Marino,$^{2,3}$
F. D'Antona$^{6}$
G. Cordoni,$^{2}$
M. V. Legnardi,$^{1}$
\newauthor
E. P. Lagioia,$^{1}$
S. Jang,$^{4}$
T. Ziliotto,$^{1}$
M. Carlos,$^{5}$
F. Dell'Agli,$^{6}$
A. Karakas,$^{7,8}$
A. Mohandasan,$^{1}$
\newauthor
Z. Osborn$^{7.8}$
M. Tailo,$^{9}$
P. Ventura$^{6}$
\\
$^{1}$ Dipartimento di Fisica e Astronomia ``Galileo Galilei'', Universit\`{a} di Padova, Vicolo dell'Osservatorio 3, I-35122, Padua, Italy\\
$^{2}$ Istituto Nazionale di Astrofisica - Osservatorio Astronomico di Padova, Vicolo dell’Osservatorio 5, Padova, IT-35122\\
$^{3}$ Istituto Nazionale di Astrofisica - Osservatorio Astrofisico di Arcetri, Largo Enrico Fermi, 5, Firenze, IT-50125\\
$^{4}$ Center for Galaxy Evolution Research and Department of Astronomy, Yonsei University, Seoul 03722, Korea\\
$^{5}$ Theoretical Astrophysics, Department of Physics and Astronomy, Uppsala University, Box 516, SE-751 20 Uppsala, Sweden\\
$^{6}$ Istituto Nazionale di Astrofisica - Osservatorio Astronomico di Roma, Via Frascati 33, I-00040 Monteporzio Catone, Roma, Italy\\
$^{7}$ School of Physics and Astronomy, Monash University, VIC 3800, Australia\\
$^{8}$ Centre of Excellence for Astrophysics in Three Dimensions (ASTRO-3D), Melbourne, Victoria, Australia\\
$^{9}$ Dipartimento di Fisica e Astronomia Augusto Righi, Universit`a degli Studi di Bologna, Via Gobetti 93/2, 40129, Bologna, Italy\\
}

\date{Accepted XXX. Received YYY; in original form ZZZ}

\pubyear{2023}

\begin{document}
\label{firstpage}
\pagerange{\pageref{firstpage}--\pageref{lastpage}}
\maketitle

\begin{abstract}

About one-fifth of the Galactic globular clusters (GCs), dubbed Type II\,GCs, host distinct stellar populations with different heavy elements abundances. NGC\,1851 is one of the most studied Type II GCs, surrounded by several controversies regarding the spatial distribution of its populations and the presence of star-to-star [Fe/H], C$+$N$+$O, and age differences. This paper provides a detailed characterization of its stellar populations through {\it{Hubble Space Telescope (HST)}}, ground-based, and Gaia photometry.
We identified two distinct populations with different abundances of s-process elements along the red-giant branch (RGB) and the sub-giant branch (SGB) and detected two sub-populations among both s-poor (canonical) and s-rich (anomalous) stars. To constrain the chemical composition of these stellar populations, we compared observed and simulated colors of stars with different abundances of He, C, N, and O. It results that the anomalous population has a higher CNO overall abundance compared to the canonical population and that both host stars with different light-element abundances.
No significant differences in radial segregation between canonical and anomalous stars are detected, while we find that among their sub-populations, the two most chemical extremes are more centrally concentrated. Anomalous and canonical stars show different 2D spatial distributions outside $\sim$3 arcmin, with the latter developing an elliptical shape and a stellar overdensity in the northeast direction.
We confirm the presence of a stellar halo up to $\sim$80 arcmin with Gaia photometry, tagging 14 and five of its stars as canonical and anomalous, respectively, finding a lack of the latter in the south/southeast field.

\end{abstract}




\section{Introduction}

Globular Clusters (GCs) host distinct groups of stars with different chemical compositions, as well established by the past few decades of research in stellar astrophysics. The multiple populations phenomenon, i.e., the evidence of star-to-star abundance variations of light elements (e.g., He, C, N, O, Al, Na), is widespread among Galactic GCs and has been detected among star clusters in the nearby galaxies, such as the Magellanic Clouds, Fornax, and M\,31. Despite the intense effort put in throughout the years, its origin is still not clear \citep[see][for reviews]{bastian2018, gratton2019, milone2022}.

An additional challenge in the field is the presence of a subset ($\sim$18\%) of GCs which, beyond the typical light-elements variations, also show the following three observational features: (i) a split sub-giant branch (SGB) in color-magnitude diagrams (CMDs) constructed with optical filters, (ii) a secondary red-giant branch (RGB) sequence, which is associated with the faint SGB, (iii) abundance variations in C$+$N$+$O, metallicity, and/or $s$-process elements \citep[see][]{milone2017}. These were defined as Type II GCs by Milone and collaborators, in opposition to the typical Milky Way Type I clusters.
Similarly to Type I GCs, Type II GCs host stellar populations with different abundances in light elements, but they also exhibit additional sequences of stars in the CMD that produce the three aforementioned features. For that, we will refer hereafter as 'canonical' the stars that yield the multiple population patterns observed in all GCs, and as 'anomalous' the stars present in Type II GCs only.

Several questions arise at this point. What is the origin of anomalous stars? Why do they appear in some GCs and not in others? Did these GCs originate through different mechanisms with respect to the typical Type I clusters? Which is the sequence of events in the star formation history of these objects that led to such a complex chemical pattern?

In this context, NGC\,1851 is one of the most intriguing and controversial Type II GCs, with numerous studies in both photometry and spectroscopy aimed to shed light on the mechanisms that produced the cluster we nowadays observe. 
Photometry was instrumental in the first discovery of anomalous features in NGC1851, with the detection of a split SGB in optical filters \citep{milone2008, milone2009, zoccali2009}. The faint and bright SGBs evolve into red and blue RGBs, respectively, clearly visible in CMDs constructed with the $U-I$ colors \citep{han2009, milone2017, jang2022}.

Spectroscopy shows that the faint SGB and the red RGB are populated by stars with enhanced abundances of s-process elements, with respect to the bright SGB and the blue RGB \citep[e.g.][]{yong2008, villanova2010, carretta2011, gratton2012, marino2014, mckenzie2022, taut2022} and that both s-rich and s-poor stars exhibit internal variations in some light elements, including C, N, O, and Na \citep[e.g.][]{yong2009, yong2015, lardo2012, carretta2010, campbell2012, carretta2014, milone2017, simpson2017, jang2022}.

The physical reasons that are responsible for the split SGB are widely debated. Works based on the comparison between photometry and stellar models reveal that the faint SGB is composed of stars that are either older by $\sim$1\,Gyr than the bright SGB or have nearly the same age as bright SGB stars but are enhanced in their overall C$+$N$+$O content by a factor of $\sim$3 \citep{cassisi2008, ventura2009, dantona2009}.  
However, spectroscopic investigations provided controversial results. \citet{yong2009, yong2015} and \citet{simpson2017} detected large differences in the overall C$+$N$+$O content of s-rich and s-poor stars, whereas other authors concluded that all stars in NGC\,1851 share the same C$+$N$+$O content \citep[e.g.\,][]{villanova2010, taut2022}.  The correct chemical characterization of multiple populations in NGC\,1851 and their relative ages is further challenged by the possibility that $s$-rich stars are enhanced in iron by $\sim$0.05-0.10 dex with respect to $s$-poor stars \citep[see][for discussion on the presence or lack of metallicity difference between $s$-rich and $s$-poor stars in NGC\,1851]{gratton2012, lardo2012, taut2022}.

Controversial conclusions come also from the radial distribution of stellar populations in NGC\,1851.
As an example, \citet{zoccali2009} concluded that the faint-SGB stars are centrally concentrated and tend to disappear moving away from the GC center. Conversely, \citet{milone2009} find a nearly constant ratio between the number of stars in the two SGBs \citep[see also][]{cummings2014}.

Overall, several unsolved issues affect our current understanding of the processes originating the complex observational features of NGC\,1851, among all the presence of an anomalous stellar population. Therefore, an accurate definition of the chemical, spatial, and kinematic properties of these stars is mandatory to explain how anomalous stars were born.

In this work, we analyze photometry from different space- and ground-based telescopes to provide new tools for the photometrical tagging of the different populations that inhabit NGC\,1851, with a particular focus on their spatial behavior.
Section~\ref{sec:data} describes the dataset used in our work. Section~\ref{sec:rgb} illustrates the method adopted to disentangle the multiple populations of NGC\,1851 among RGB stars, while their chemical composition is inferred in Section~\ref{sec:chimica}.
Section~\ref{sec:lower} is dedicated to multiple populations along the SGB and the MS. Section~\ref{sec:fraction} presents the calculation of the fractions of the multiple stellar populations spotted in this cluster and explores their radial distribution, while the 2D spatial distribution of canonical and anomalous stars is investigated in Section~\ref{sec:spatial}. Finally, Section~\ref{sec:final} provides a summary and conclusions.

\section{Dataset} \label{sec:data}

In this work, we exploit three photometric datasets. First, we build a catalog of the innermost $\sim$2.7$\times$2.7 arcmin$^{2}$ stars by exploiting {\it{Hubble Space Telescope (HST)}} observations taken with the Ultraviolet and Visual Channel of the Wide Field Camera 3 (WFC3/UVIS) filters F275W, F336W, and F438W (GO-13297), and the Wide Field Channel of the Advanced Camera for Survey (ACS/WFC) filters F606W and F814W (GO-10775).
We perform effective point-spread function (PSF) photometry \citep[see][]{anderson2000} to obtain accurate stellar positions and magnitudes through the KS2 software, developed by Jay Anderson \citep[see][for details]{sabbi2016a, bellini2017a, milone2023a}, which is an extended version of the program kitchen\_sync \citep{anderson2008}.
To the obtained catalog, we apply the quality diagnostics described in \citet[][see their Sections 2.5 and 3 for details]{nardiello2018} to select stellar sources with high-quality astrometry and photometry.
No correction for differential reddening has been performed since this cluster is characterized by very small reddening variations \citep[e.g.,][]{jang2022, legnardi2023}, which produce negligible effects on the photometric quality of the catalog. We instead correct this catalog for zero-point spatial variations effects, following the recipe presented in \citet[][see their Section 3.2]{milone2012c}.

\begin{figure*}
    \centering
    \includegraphics[trim={0 0 0 0},clip,width=14cm]{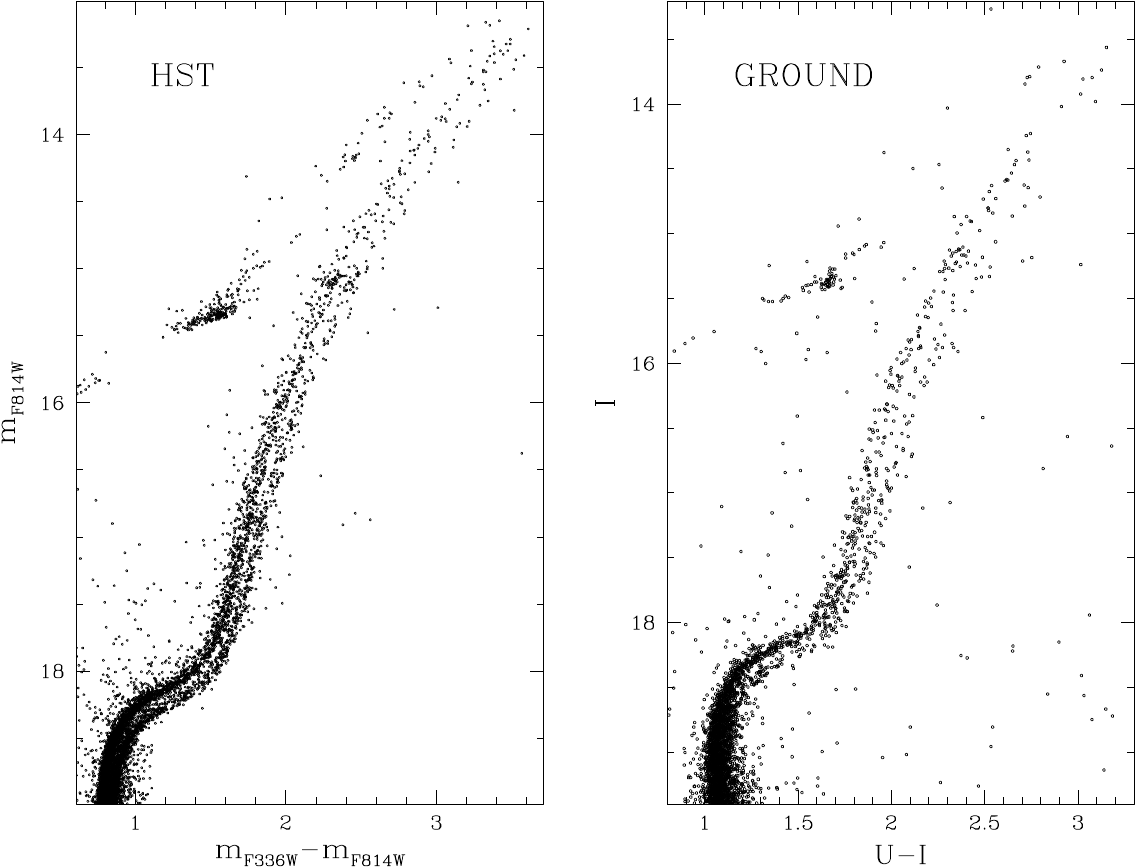}
    \caption{{\it{Left panel:}} $m_{\rm F814W}$ vs. $m_{\rm F336W}$-$m_{\rm F814W}$ CMD obtained from {\it HST} photometry. {\it{Right panel:}} $I$ vs. $U$-$I$ CMD obtained from the ground-based observations. 
    }
    \label{fig:cmd}
\end{figure*}

To investigate the cluster regions outside the {\it HST} Field of view (FoV), we use the ground-based catalog by \citet{stetson2019}. This catalog includes stellar magnitudes in the $U$, $B$, $V$, $R$, and $I$ bands and reaches distances from the cluster's center up to $\sim$20 arcmin. It was built by performing PSF photometry on images from multiple ground-based facilities taken at different epochs. To this catalog, we apply a cleaning procedure to isolate the cluster's stars through the diagnostics defined by Stetson and collaborators (see their Section 4.1), and a correction for zero-points variations by extending the procedure used on the {\it HST} catalog to this dataset \citep[see also][]{jang2022}.

Figure~\ref{fig:cmd} presents examples of the resulting {\it HST} and ground-based CMDs. Specifically, we show the $m_{\rm F814W}$ vs. $m_{\rm F336W}$-$m_{\rm F814W}$ CMD (left panel) from {\it HST} photometry, and the $I$ vs. $U$-$I$ (right panel) CMD from ground-based photometry. In both diagrams, two sequences are clearly distinguishable from the SGB up to the RGB tip. 

Finally, we exploit Gaia Data Release 3 \citep[DR3, ][]{Gaia2021} observations to explore the stars in the halo of NGC\,1851 (i.e., at distances much larger than the tidal radius), reaching a radial distance of about 80 arcmin from the center. This feature will be discussed in Section~\ref{sec:spatial}.

\subsection{Artificial star test} \label{sec:as}

We perform artificial-star (AS) tests to estimate the photometric errors in the {\it HST} dataset and to account for the effects of the large crowding in the innermost regions.
To do that, we applied the procedure described in \citet{anderson2008}, which consists in adding ASs (i.e., sources with known position and magnitude) into the images and then applying to them the reduction procedure used on real stars.

Each test performed in this work is based on a catalog of 100,000 ASs, which position and magnitude are defined by following the crowding distribution and the CMDs sequences described by the observed stars, respectively.
To pass the test stars must have position and magnitude differences smaller than 0.5 pixel and 0.75 mag, respectively, between this input catalog and the output produced after the data reduction.
These stars are then used to estimate the photometric errors and the amount of contamination that a given stellar population introduces in the area of photometric diagrams belonging to another stellar population, and hence the uncertainties associated with the population ratios inferred in Section~\ref{sec:fraction}.

\section{A zoo of populations along the Red Giant Branch}\label{sec:rgb}

\begin{figure*}
    \centering
    \includegraphics[trim={0 -0.2cm 0 0},clip,height=7.7cm]{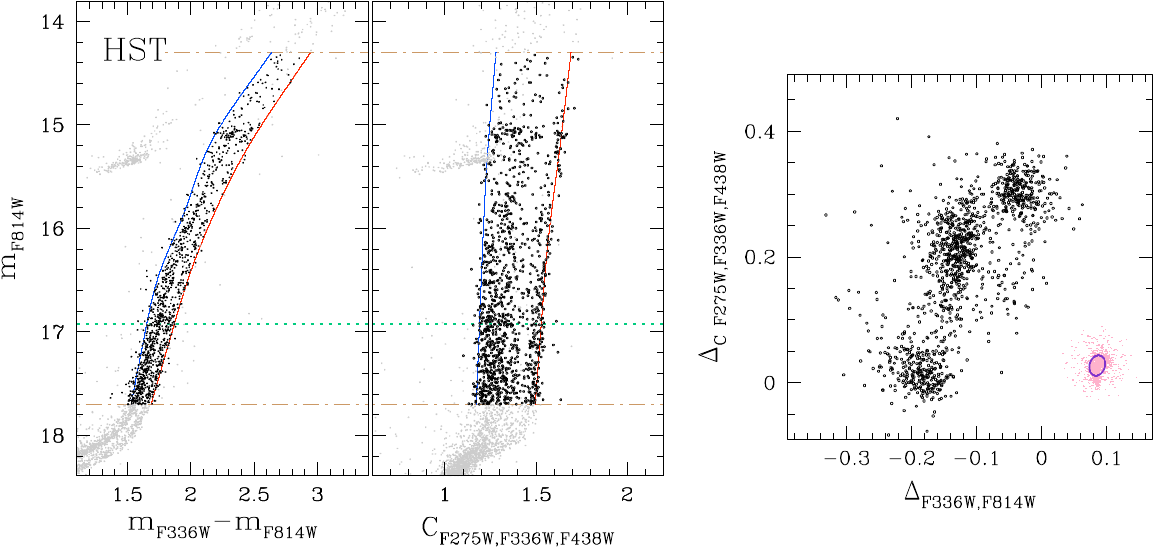}
    \includegraphics[trim={0 0 0 -0.2cm},clip,height=7.7cm]{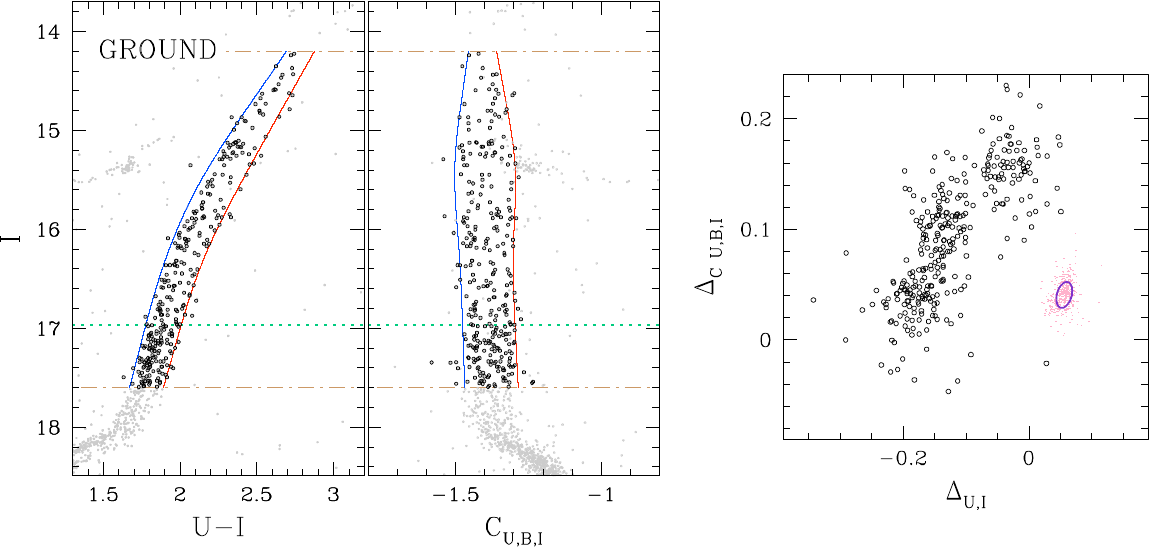}
    \caption{{\it{Top-left and -middle panels:}} $m_{F814W}$ vs.\,$m_{\rm F336W}$-$m_{F814W}$ CMD and $m_{F814W}$ vs.\,$C_{\rm F275W,F336W,F438W}$ pseudo-CMD of stars in the {\it HST} FoV.  {\it{Top-right panel:}} $\Delta_{\rm C F275W,F336W,F438W}$ vs.\,$\Delta_{\rm F336W,F814W}$ ChM of RGB stars.
    {\it{Bottom-Left and -middle panels:}} $I$ vs. $U$-$I$ CMD and $I$ vs. $C_{\rm U,B,I}$ pseudo-CMD of stars in the ground field. 
    {\it{Bottom-right panel:}} $\Delta_{\rm C U,B,I}$ vs. $\Delta_{\rm U,I}$ ChM of RGB stars. The brown dot-dashed horizontal lines separate the stars included (black points) and excluded (grey points) from each ChM determination. The dotted aqua lines indicate the magnitude level at which the ChM widths were normalized (see the text for details).
    Pink points illustrate the distribution in both ChMs of a simulated single stellar population, while the purple ellipses include 68.27\% of the simulated stars.
    }
    \label{fig:pchm}
\end{figure*}

In this Section, we explore {\it HST} and ground-based photometry of RGB stars to identify the multiple populations of NGC\,1851. To do this, we adopt the Chromosome Map (ChM), which is a two pseudo-color diagram that maximizes the separation between chemically-different populations \citep[see][for details]{milone2015, milone2017}. We introduce two new ChMs that maximize at the same time the separations of the canonical and anomalous stellar populations and of the populations with different light-element abundances.

The $m_{\rm F336W}$-$m_{\rm F814W}$ and the analogous $U$-$I$ color are effective tools to separate the blue and red RGBs that host the canonical and anomalous stars, respectively \citep[see also][]{han2009, milone2017}.
In the {\it HST} dataset, we combine this information with the $C_{\rm F275W,F336W,F438W} =$ $m_{\rm F275W}$-2$m_{\rm F336W}$+$m_{\rm F438W}$, which is sensitive to stellar populations with different carbon, nitrogen, and oxygen content \citep[e.g.][and references therein]{milone2022}.

The procedure to derive this ChM is illustrated in Figure~\ref{fig:pchm} for RGB stars with $14.3< m_{\rm F814W} < 17.7$ (black dots), where the separation between different sequences is well visible in both filter combinations.
We follow the recipe by \citet[][see their Section 3.1 and 3.2]{milone2017} to derive the red and blue boundaries of both RGBs. Moreover, we calculated the RGB widths, defined as the difference between the red and blue boundaries at a magnitude level of 2 $m_{\rm F814W}$ above the MS Turn-Off (dotted aqua line). 
Finally, by applying their Equations (1) and (2), we derive the ChM coordinates $\Delta_{\rm F336W,F814W}$ and $\Delta_{\rm C F275W,F336W,F438W}$, plotted in the top-right panel.
We used the AS photometry to simulate a single stellar population in the ChM plane.
The simulated points are arbitrarily shifted near the bottom-right corner of the ChM and represented in pink, while the purple ellipse includes 68.27\% of them.

\begin{figure*}
    \centering
    \includegraphics[trim={0 0 -0.4cm 0},clip,width=7.0cm]{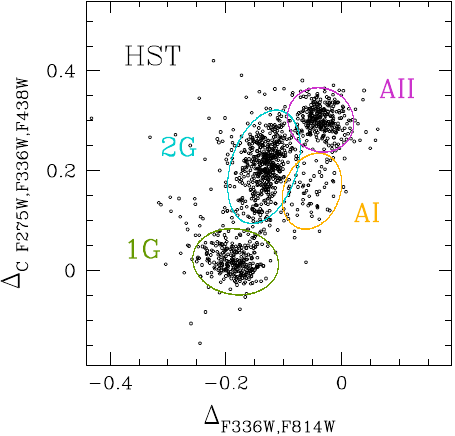}
    \includegraphics[trim={-0.4cm 0 0 0},clip,width=7.0cm]{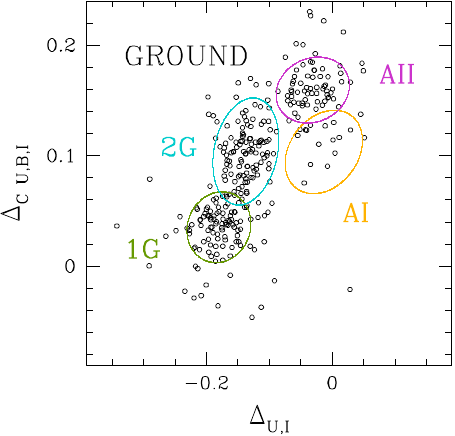}
    \caption{Elliptical regions that encapsulate each spotted population in the {\it HST} and ground-based (left and right panels, respectively) ChMs. Green and azure ellipses define the 1G and 2G regions of canonical stars, while the yellow and purple ones are the AI and AII regions of anomalous stars.
    }
    \label{fig:ellipse}
\end{figure*}

Similarly, we derived the ChM from ground-based photometry, by using the $U$-$I$ color, which is analogous to $m_{\rm F336W}$-$m_{\rm F814W}$, and $C_{\rm U,B,I} =$ $U$-2$B$+$I$ pseudo-color, which is an efficient tool to separate stellar populations with different light-element abundances \citep[e.g.][]{jang2022}.
The $I$ vs. $U$-$I$ and the $I$ vs. $C_{\rm U,B,I}$ diagrams, and the resulting ChM are shown in the bottom panels of Figure~\ref{fig:pchm}.

As illustrated in Figure\,\ref{fig:pchm}, the canonical and anomalous stars define two distinct sequences in both ChMs with  $\Delta_{\rm F336W,F814W}$ (or $\Delta_{\rm U,I}$) smaller and larger than $\sim -0.1$, respectively.
Both canonical and anomalous stars show $\Delta_{\rm C F275W,F336W,F438W}$ and $\Delta_{\rm C U,B,I}$ distributions wider than what expected by observational errors only.
This fact demonstrates that both RGBs present variations in their light-elements abundances.
Specifically, we detect the first- and second-population stars typically present in GCs (hereafter 1G and 2G) along the canonical RGB, forming two separate blobs in both ChMs, and two anomalous populations distinguishable in the {\it HST}-based ChM (hereafter AI and AII).
Intriguingly, 2G stars span wider intervals of $\Delta_{\rm C F275W,F336W,F438W}$ and $\Delta_{\rm C U,B,I}$ than the other populations, thus indicating that their stars are not chemically homogeneous. 
 
The ChM regions occupied by the bulk of 1G, 2G, AI, and AII stars are enclosed in the ellipses displayed in Figure~\ref{fig:ellipse}. 
These ellipses, defined as in \citet{dondoglio2022}, are used to estimate the fraction of stars in each population.
In a nutshell, we first select by hand the bonafide members of each population and measure their median ChM coordinates to define the center of the ellipse. Secondly, to find the major axis direction, we consider the direction of a line that crosses the center and minimizes the orthogonal dispersion of the bonafide members. Finally, we fix the length of the semi-major and -minor axis as 2.5 times the dispersion of stars along the directions parallel and orthogonal to the major axis direction, respectively.
We show encircled in green and azure 1G and 2G stars, respectively, and in yellow and purple the two groups of anomalous stars, AI and AII.
Due to the small number of stars, the AI stars do not form a clear distinguishable blob in the ground-based ChM. Although we could not classify them with confidence as a distinct population by using this diagram alone, we define by eye an ellipse that encloses the probable AI stars identified from ground-based photometry.

The fraction of stars in each population is calculated  as in previous work from our group. As an example, to estimate the fraction of 1G stars, we first derived the number of stars within the green ellipse, which provides a very crude estimate of the number of 1G stars. 
To estimate the total number of 1G stars we first subtracted to this value the numbers of 2G, AI, and AII stars that, due to observational errors, lie within the green ellipse. Then, we added the number of 1G stars outside the green ellipse.  Similarly, we derived the fraction of 2G, AI, and AII stars.
The fraction of stars of each population within the four ellipses is inferred by means of ASs \citep[see][for details]{milone2012a, zennaro2019a, dondoglio2022}. 
    
The results are summarized in Table\,\ref{tab:fraction}. We find that $\sim$70\% and $\sim$30\% of stars belong to the canonical and anomalous population, respectively. 1G stars comprise more than one-third of the total number of canonical stars, whereas AI stars include less than 10\% of the anomalous stars.

\section{The chemical composition of the multiple populations in NGC\,1851}\label{sec:chimica}

\begin{figure*}
    \centering
    \includegraphics[trim={0 0 0 0},clip,width=18cm]{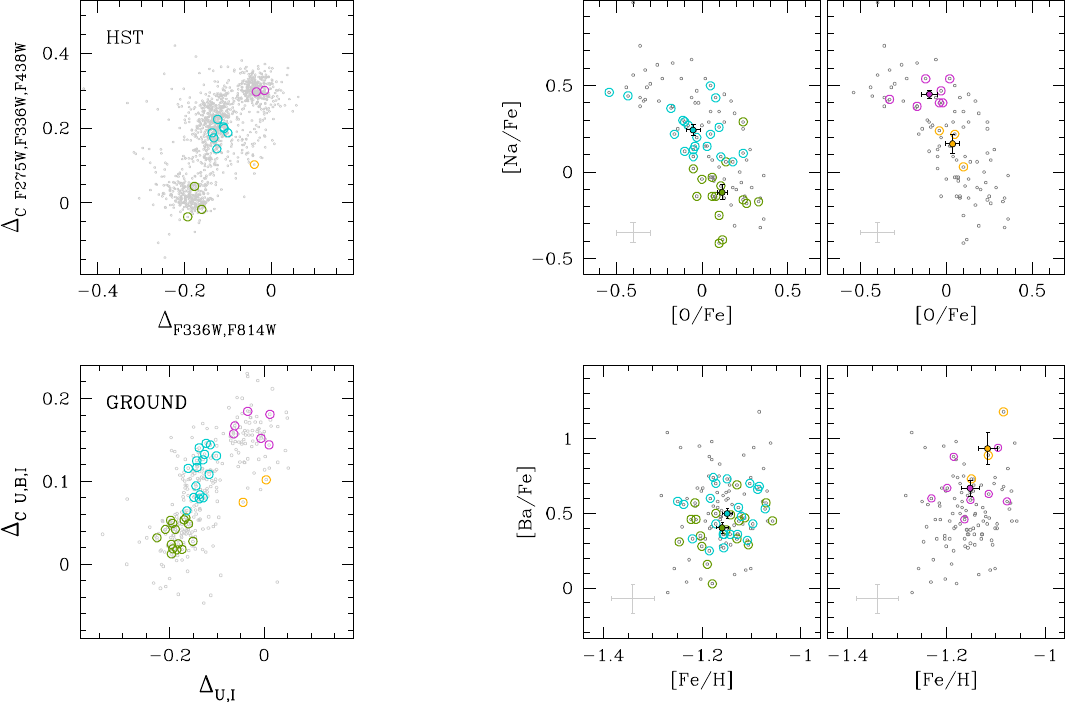}
    \caption{ 
    {\it{Left panels:}} HST (upper) and ground-based (bottom) ChMs where the stars in common with the spectroscopic datset from \citet{carretta2011} are highlighted with open bullets, color-coded following the prescriptions of Figure~\ref{fig:ellipse}.
    {\it{Middle panels:}} Reproduction of the [Na/Fe] vs.\,[O/Fe] and [Ba/Fe] vs.\,[Fe/H] relations for the two canonical populations (upper and lower panels, respectively). Dark-grey points represent all the stars in the Carretta and collaborators dataset. Filled dots with black contours mark the average abundance of stars in each population and black bars indicate their errors. Gray bars highlight the average uncertainties of the spectroscopic measurements.
    {\it{Right panels:}} same as the middle panels but for the two anomalous populations.
    }
    \label{fig:spectra}
\end{figure*}

To derive the average chemical compositions of the four stellar populations that we identified along the RGB, we combine information from photometry and spectroscopy. The left panels of Figure~\ref{fig:spectra} show the ChMs introduced in Figure~\ref{fig:pchm}, where we encircle stars in common with the spectroscopic dataset by \citet{carretta2011} from GIRAFFE spectra of 124 RGB stars, color-coded according to their belonging to the ellipses defined in Figure~\ref{fig:ellipse}.
The upper-middle and -right panels display the sodium-oxygen anticorrelation among the canonical and anomalous stars for which both our photometric tagging and abundance measurements from Carretta and collaborators are available.
Filled points with black contours indicate the average abundances\footnote{
Other spectroscopic datasets with similar information \citep[e.g.,][]{meszaros2020, taut2022} were not considered in this comparison, because the number of stars with ChMs tagging in the considered magnitude intervals is much lower than the Carretta and collaborators' catalog, not allowing us for meaningful estimates of the average abundances of the four populations.
}.
The anomalous stars span a smaller range of [Na/Fe] and [O/Fe] with respect to canonical ones. As expected, we find that 2G stars are enhanced in sodium and depleted in oxygen, with respect to the 1G. Similarly, AII stars are more sodium-rich and oxygen-poor than AI stars. 
Intriguingly, AI stars exhibit higher sodium content than 1G stars in close analogy with what is observed in other Type\,II GCs \citep{marino2009a, marino2011a, marino2011b}.

The lower-middle and -right panels of Figure\,\ref{fig:spectra} show that the anomalous stars are enhanced in [Ba/Fe] with respect to canonical stars, while their average [Fe/H] are consistent within uncertainties. There is no evidence for internal variations among canonical stars, while AI have larger average [Ba/Fe] and [Fe/H] than AII stars (even though our sample includes only three AI stars).
Table~\ref{tab:fraction} reports the average [O/Fe], [Na/Fe], [Ba/Fe], and [Fe/H] of canonical and anomalous stars and their subpopulations. 

Spectroscopic results corroborate the conclusions inferred from photometry alone. The finding of distinct groups of 1G-2G and AI-AII within the canonical and anomalous RGB of NGC\,1851 are in agreement with the presence of Na-O anti-correlation in both RGBs \citep[e.g.,][]{carretta2011, taut2022}. We also note that the three AI stars with available spectroscopy, including two AI stars selected from ground-based photometry, are more Na-poor than AII stars.  
This fact supports our choice of the elliptical regions used to select AI stars in the {\it HST} and ground-based ChMs. 

\begin{figure*}
    \centering
    \includegraphics[trim={3.5cm 7cm 8.5cm 12.7cm},clip,height=9cm]{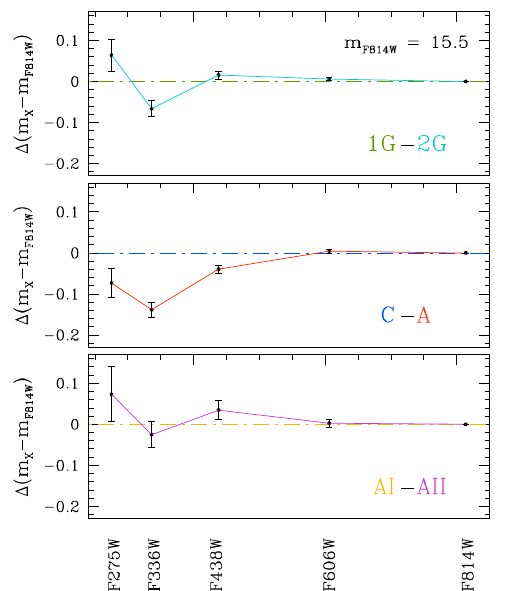}
    \includegraphics[trim={3.0cm 7cm 9cm 12.7cm},clip,height=9cm]{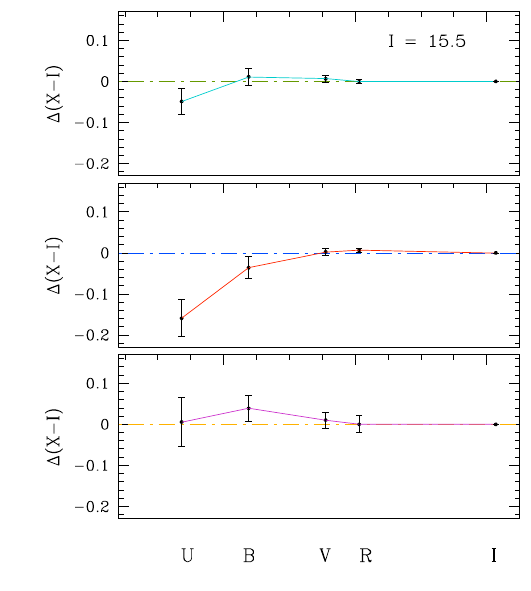}
    \caption{ {\it{Left panels:}} $\Delta (m_{\rm X}-m_{\rm F814W})$ between different populations in the {\it HST} filters at a magnitude level $m_{\rm F814W}$ 15.5, with X$=$F275W, F336W, F438W, F606W, and F814W. From top to bottom, 1G and 2G, canonical and anomalous, AI and AII stellar populations are compared.
    {\it{Right panels:}} same as left panels but for the ground-based filters. Here, X$=$U, B, V, R, and I.
    }
    \label{fig:rdmag}
\end{figure*}

To further investigate the chemical composition of the stellar populations in NGC\,1851, we combine information from multi-band photometry and synthetic spectra.
To do that, we apply to our dataset a method widely used in previous papers from our group \citep[e.g.][]{milone2012b, lagioia2019a}, which allows us to constrain the relative abundances of helium, carbon, nitrogen, and oxygen of two stellar populations. 
Specifically, we compared the chemical composition of 2G and 1G stars, AII and AI stars, and anomalous and canonical stars.
In a nutshell, we first derive fiducial lines along the RGB for each population in the $m_{\rm F814W}$ vs.\,$m_{\rm X}-m_{\rm F814W}$ (for {\it HST} observations) and in the $I$ vs. $X-I$ (for ground-based data) CMDs, with X$=$F275W, F336W, F438W, F606W, and F814W, and X$=$U, B, V, R, and I, respectively.
Then, we identify three equally-spaced reference magnitudes ($m_{\rm ref}$) fainter than the RGB bump.
For each value of $m_{\rm ref}$ we measure the color differences ($\Delta(m_{\rm X}-m_{\rm F814W})$ and $\Delta (X-I)$) between the two population fiducial lines.  We portray in Figure~\ref{fig:rdmag} the relative colors at $m_{\rm ref}=15.5$ mag.

Qualitatively, the color differences between 2G and 1G stars for the different filters (upper panels) follow a similar pattern to what is generally observed in most GCs \citep[e.g.][]{milone2018}. The 2G stars are typically bluer than the 1G, and their color separation reaches its maximum when using a wide color baseline such as $m_{\rm F275W}-m_{\rm F814W}$. The F336W/U band provides a remarkable exception because the 2G stars exhibit redder $m_{\rm F336W}-m_{\rm F814W}$ (and $U-I$) colors than the 1G.

To infer the relative abundances of 2G and 1G stars, we first derive the values of the effective temperature ($T_{\rm eff}$) and gravity ($g$) corresponding to each value of $m_{\rm ref}$ by using the best-fitting isochrones from \citet{ventura2009} and \citet{dantona2009}.
We compute a reference spectrum, with pristine helium content of $Y$=0.246, [O/Fe]=0.4 dex, solar carbon abundance, and [N/Fe]=0.5 dex. Moreover, we derive a grid of comparison spectra with helium mass fractions ranging from Y=0.246 to 0.280 in steps of 0.001, [O/Fe] from 0.0 to 0.6 in steps of 0.1 dex, while both [C/Fe] and [N/Fe] span the intervals between $-0.5$ to 0.2 dex and between 0.5 and 2.0 dex, respectively in steps of 0.1 dex. 
When we used the He-enhanced chemical composition, we adopted the corresponding values for the effective temperature and gravity derived by the isochrones.
The spectra are computed by using the ATLAS12 and SYNTHE computer programs \citep[e.g.][]{castelli2005a, kurucz2005a, sbordone2007a}.
We used isochrones with constant C$+$N$+$O abundance.
We find that 2G stars are enhanced in nitrogen by 0.80$\pm$0.10 dex, and depleted in carbon and oxygen by 0.25$\pm$0.10 and 0.20$\pm$0.10 dex respectively, when compared to 1G stars. Moreover, they have a slightly larger helium mass fraction ($\Delta Y$=0.008$\pm$0.006) than the 1G. 
The errors are estimated as the dispersion of the abundance determinations corresponding to the three magnitude levels, divided by the square root of two.
We repeated the same analysis by using isochrones from the Dartmouth database \citep{dotter2008} and obtained similar conclusions. Specifically, we inferred differences in [C/Fe], [N/Fe], and [O/Fe] between 2G and 1G stars of 0.85$\pm$0.10 dex, and depleted in carbon and oxygen of $-$0.25$\pm$0.10 and $-$0.30$\pm$0.10 dex, respectively. Moreover, we find a difference in helium mass fraction of ($\Delta Y$=0.007$\pm$0.005).

The relative colors of canonical and anomalous stars (middle panels) in different filters differ significantly from those of 1G and 2G stars. 
The anomalous RGB exhibits redder $m_{\rm X}-m_{\rm F814W}$ ($X-I$) colors than the canonical RGB for X=F275W, F336W, and F438W (X=U and B) with similar F814W (I) magnitudes. Conversely, the color difference disappears in the F606W$-$F814W, $V-I$, and $R-I$ colors.

By assuming the same overall C$+$N$+$O content and helium content for the canonical and the anomalous stars, we find that the latter would be enhanced in C and O by $\sim$0.9 and $\sim$0.8 dex, respectively, and  in N by $\sim$0.9 dex compared to the canonical population.
Noticeably, these results on carbon and oxygen would be in disagreement with the conclusions of papers based on high-resolution spectroscopy. As an example, both \citet{yong2015} and \citet{taut2022} find nearly the same values of [C/Fe] for the canonical and anomalous population, with the latter being slightly depleted in oxygen by $\sim$0.2 dex with respect to the canonical population. 
Moreover, according to the results that we inferred from multi-band photometry, the anomalous stars would be significatively enhanced in their overall C$+$N$+$O abundance. This fact would indicate that the atmospheric parameters that we used to compute the spectra of anomalous stars, which are derived from isochrones with constant C$+$N$+$O content, are not correct.

To further investigate the chemical composition of anomalous and canonical stars, we estimate the relative abundances of anomalous and canonical stars by using the method above and the atmospheric parameters inferred from the isochrones by \citet{dantona2009} and \citet{ventura2009}. These isochrones have different C$+$N$+$O content, pristine helium abundance Y=0.25, and reproduce the double SGB of NGC\,1851.
The most remarkable difference with the isochrones with constant C$+$N$+$O abundance is that, for a fixed F814W magnitude, the RGB of CNO-enhanced stars is colder by $\sim 30$K than the canonical RGB.

By assuming that the canonical and anomalous stars share the same helium content, we reproduce their relative colors by assuming that the anomalous are enhanced in nitrogen by 0.90$\pm$0.15 dex and share the same carbon and oxygen abundances  $\Delta$[C/Fe]=0.10$\pm$0.15, $\Delta$[O/Fe]=$-$0.05$\pm$0.15). We thus confirm the results by \citet{yong2015} based on high-resolution spectroscopy.
 
Finally, we infer the relative abundances of AI and AII stars (lower panels), by using the same approach used for the 1G and 2G stars. Based on the isochrones from the Roma database \citep{ventura2009, dantona2009}, we find that AII stars have slightly higher content of helium and nitrogen ($\Delta$Y=0.005$\pm$0.013 and $\Delta$[N/Fe]=0.30$\pm$0.20 dex), and lower abundances of carbon and oxygen ($\Delta$[C/Fe]=$-$0.25$\pm$0.15 and $\Delta$[O/Fe]=$-$0.20$\pm$0.10 dex). We obtain similar conclusions by using the ischrones from \citet{dotter2008} ($\Delta$Y=0.006$\pm$0.011, $\Delta$[C/Fe]=$-$0.30$\pm$0.15, $\Delta$[N/Fe]=0.40$\pm$0.20, and $\Delta$[O/Fe]=$-$0.15$\pm$0.15 dex).

\subsection{Comparison with Yong et al. (2015)}

\begin{figure*}
    \centering
    \includegraphics[trim={0 0 0 0},clip,height=11cm]{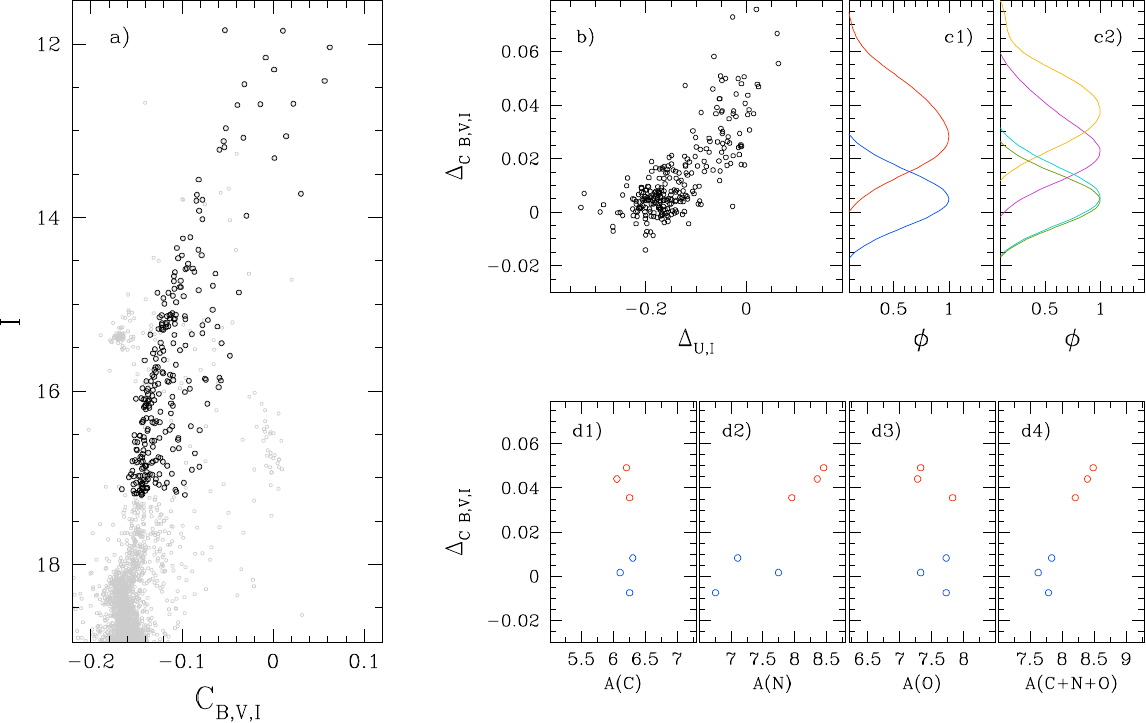}
    \caption{ {\it{Panel a):}} $I$ vs. $C_{\rm B,V,I}$ pseudo-CMD. RGB stars with 11.8$<$I$<$17.2 are highlighted with black points, while the remaining stars are colored in grey. {\it{Panel b):}} $\Delta_{\rm C B,V,I}$ vs. $\Delta_{\rm U,I}$ ChM for stars marked with black points in panel a). {\it{Panels c1) and c2):}} $\Delta_{\rm C B,V,I}$ kernel density distribution of canonical and anomalous stars and their sub-populations, respectively. {\it{Panels d1)-d4):}} $\Delta_{\rm C B,V,I}$ vs. C, N, O, and C$+$N$+$O abundances for stars in common with the \citet{yong2015} dataset. Blue and red points highlight the stars that, according to their position on the $\Delta_{\rm C B,V,I}$ vs. $\Delta_{\rm U,I}$ ChM, are canonical and anomalous, respectively.
    }
    \label{fig:cbvi}
\end{figure*}

The relative differences in the F438W/B bands can explain the $C_{\rm B,V,I}$ = $B$-2$V$+$I$ pseudo-color distribution of RGB stars in the ground-based catalog displayed in panel a) of Figure~\ref{fig:cbvi}.
This combination, introduced by \citet[][]{marino2015} and \citet{marino2019}, revealed to be particularly effective in disentangling canonical and anomalous stars in every studied Type II GC, with the latter having a wider pseudo-color distribution than the former. 
To better highlight this feature, we build the $\Delta_{\rm C B,V,I}$ vs. $\Delta_{\rm U,I}$ ChM by considering the RGB stars between 17.2<$I$<11.8 (black stars in panel a)). The result is portrayed in panel b), where canonical and anomalous stars (at $\Delta_{\rm U,I}$ smaller and bigger than $\sim-$0.1, respectively) are characterized by different extensions along the y-axis, with the latter spread over a larger $\Delta_{\rm C B,V,I}$ range, as also shown by the kernel density distributions of the two populations represented in panel c1). Panel c2) represents instead the kernel density distribution of the four populations, in which we notice that while the two canonical sub-populations are distributed almost equally, AI and AII stars are clustered around different $\Delta_{\rm C B,V,I}$, $\sim$0.02 and $\sim$0.04, respectively.
The observed behavior of NGC\,1851's stars in this color combination is consistent with the results illustrated in Figure~\ref{fig:rdmag}, so with canonical having an almost null internal spread in F438W/B and with anomalous being enhanced in these magnitudes and presenting a significant spread among their two sub-populations.

We explore the link between this pseudo-color and the C, N, and O abundances inferred by \citet{yong2015}, who measured these quantities for a sample of 15 giants and concluded that anomalous stars are enriched in total C$+$N$+$O.
Panels d1)-d4), from left to right, illustrate $\Delta_{\rm C B,V,I}$ vs. the C, N, O, and the total C$+$N$+$O abundance for the six stars from the Yong and collaborators dataset which we can characterize in the $\Delta_{\rm C B,V,I}$ vs. $\Delta_{\rm U,I}$ ChM. We colored in blue and red the stars that based on our photometric tagging belong to the canonical and anomalous population.
The C and O values of the two populations span similar ranges, suggesting no significant variations between the two in these two elements, while the N abundance correlates with $\Delta_{\rm C B,V,I}$ (Spearman’s rank correlation coefﬁcient 0.94), and becomes larger among anomalous stars.
This leads also to a correlation with the total C$+$N$+$O, corroborating our results obtained through synthetic spectra.

\section{Multiple populations along the sub-giant branch and the main sequence} \label{sec:lower}

The RGB stars are fertile ground to study multiple populations. This is in part due to the fact that they are among the brightest GC stars and typically have low photometric errors. Moreover, thanks to their structure and atmospheric parameters, the luminosity and the colors of RGB stars can be very sensitive to the abundance of some light elements. But can we identify the counterparts of the populations defined in the previous Section even among fainter stars? In this Section, we analyze the SGB and the MS to explore the populations of NGC\,1851 among these stars.

\subsection{The sub-giant branch of NGC\,1851}\label{sec:sgb}

Figure\,\ref{fig:cmd} reveals that NGC\,1851 exhibits a split SGB and that the bright and the faint SGBs are the counterparts of the canonical and anomalous RGBs, respectively.

\begin{figure*}
    \centering
    \includegraphics[clip,width=15cm]{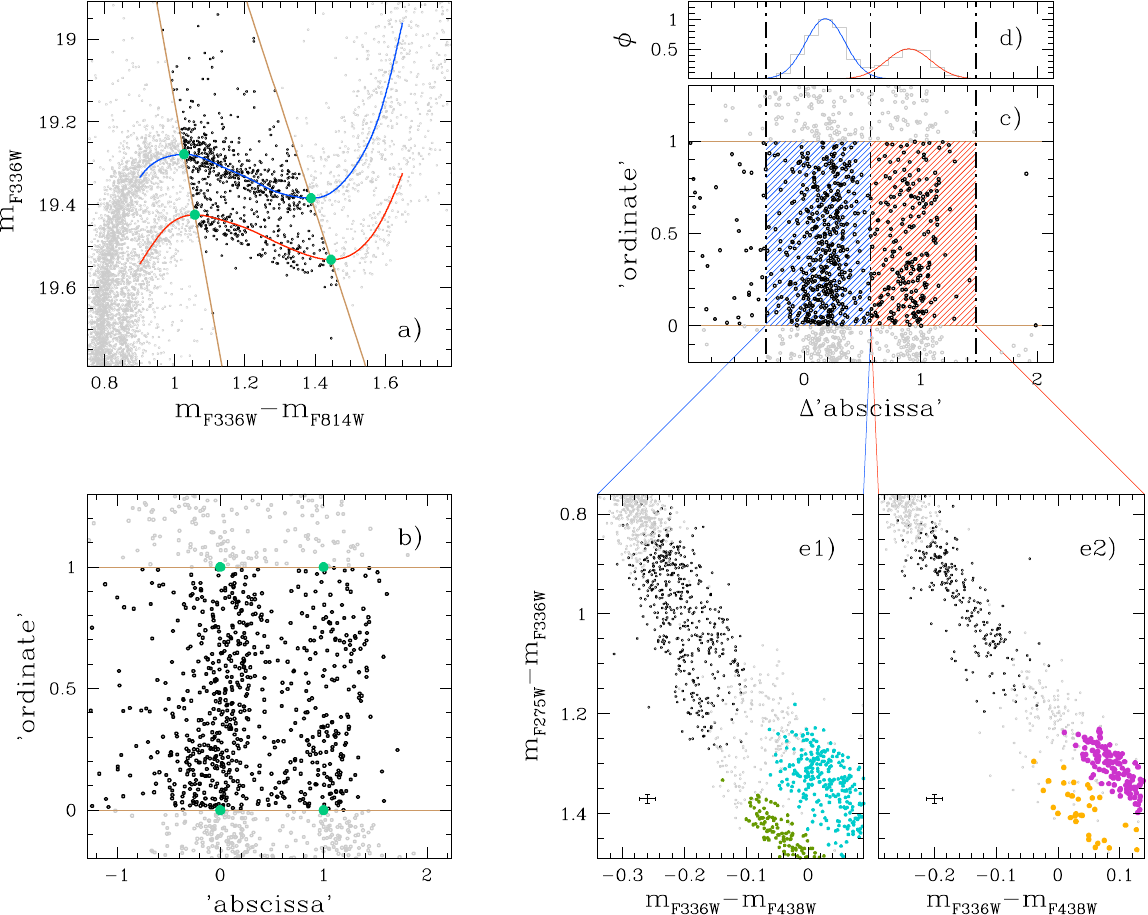}
    \caption{
    {\it{Panel a):}} $m_{\rm F336W}$ vs. $m_{\rm F336W}$-$m_{\rm F814W}$ CMD zoomed around the SGB. The blue and red lines indicate the fiducials of the canonical and anomalous SGB stars, respectively, while the aqua points represent their median brightest and faintest magnitude. The two brown lines delimit the considered SGB sample of stars. {\it{Panel b):}} 'ordinate' vs. 'abscissa' diagram of SGB stars, where lines, symbols, and colors have the same meaning as the previous panel (see text for details). {\it{Panel c):}} verticalized 'ordinate' vs. '$\Delta$abscissa' diagram, where the three vertical black dot-dashed lines delimit the region within which canonical and anomalous stars lie, colored in blue and red, respectively. {\it{Panel d):}} histogram (in grey) and best-fit Gaussian functions of the two SGB populations (colored as in panel c)).
    {\it{Panel e1) and e2):}} $m_{\rm F275W}$-$m_{\rm F336W}$ vs. $m_{\rm F336W}$-$m_{\rm F438W}$ two-color diagrams for stars inside the blue and red regions identified in panel c), respectively (black dots). Blue and red lines connect these two regions to their respective two-color diagram.
    Grey points represent the MS and RGB prosecutions of each SGB (selected on the CMD). RGB stars tagged with the ChM presented in Section~\ref{sec:rgb} are color-coded as in Figure~\ref{fig:ellipse}.
    Error bars are shown in purple.
    }
    \label{fig:sgb}
\end{figure*}

To identify the sub-populations of each SGB, we follow the procedure illustrated in Figure\,\ref{fig:sgb}.
We first select the bulk of bright and faint-SGB stars from the $m_{\rm F336W}$ vs. $m_{\rm F336W}$-$m_{\rm F814W}$ CMD displayed in panel a).
For this purpose, we derive the fiducial lines of the two SGBs by selecting bonafide stars, calculating the median color and magnitude in different color bins, and fitting these pairs of points with cubic splines (blue and red lines in panel a)).
We calculate the maximum and minimum magnitudes of both fiducials (aqua bullets) and use them to select bright- and faint-SGB stars at similar evolutionary stages (black points in Figure\,\ref{fig:sgb}a) as the ones between the two lines that cross the bluest and reddest pair of aqua points.  

To better investigate multiple populations among each SGB, we apply to the stellar colors and magnitudes the transformations by ~\citet[][see their Appendix A]{milone2009} that allow us to define the reference frame ('abscissa', 'ordinate') shown in Figure\,\ref{fig:sgb}b). In this diagram, the brown lines defined in panel a) are horizontal at 'ordinate' 0 and 1 and the four aqua points have coordinates (0,0), (0,1), (1,0), and (1,1). The canonical and anomalous SGBs form the sequences centered around 'abscissa' 0 and 1, respectively. We then apply the method described in Section~\ref{sec:rgb} to derive the 'abscissa' red and blue boundaries that we use to calculate the '$\Delta$abscissa' values plotted in panel c) against the 'ordinate'.

To improve the selection of SGB stars, we derive the histogram distribution of $\Delta$abscissa' (see panel d)) and fit it with a function given by the sum of two Gaussians by means of least squares. The two components of the best-fit function are represented in blue and red in Figure\,\ref{fig:sgb}d).
We exclude stars outside the external dot-dashed lines, which are obtained shifting the center of the blue and the red Gaussian function by three times their standard deviations. 
The central line separates the regions populated by the bulk of canonical and anomalous SGB stars and corresponds to the '$\Delta$abscissa' value at the minimum of the bi-Gaussian function. We use these lines to define the blue and red regions in panel c), which contain our sample of canonical and anomalous stars, respectively.
We exploit these regions to evaluate the fraction of canonical and anomalous SGB stars. To do that, we measure the number of stars within each interval and then we correct it by means of ASs (see Section~\ref{sec:as}) repeating the procedure applied in Section~\ref{sec:rgb}. The resulting ratios obtained by analyzing both the {\it HST} and the ground-based catalogs are listed in Table~\ref{tab:fraction} and are consistent with the values inferred from the RGB stars.

\begin{figure*}
    \centering
    \includegraphics[trim={0 0 0 0},clip,width=16cm]{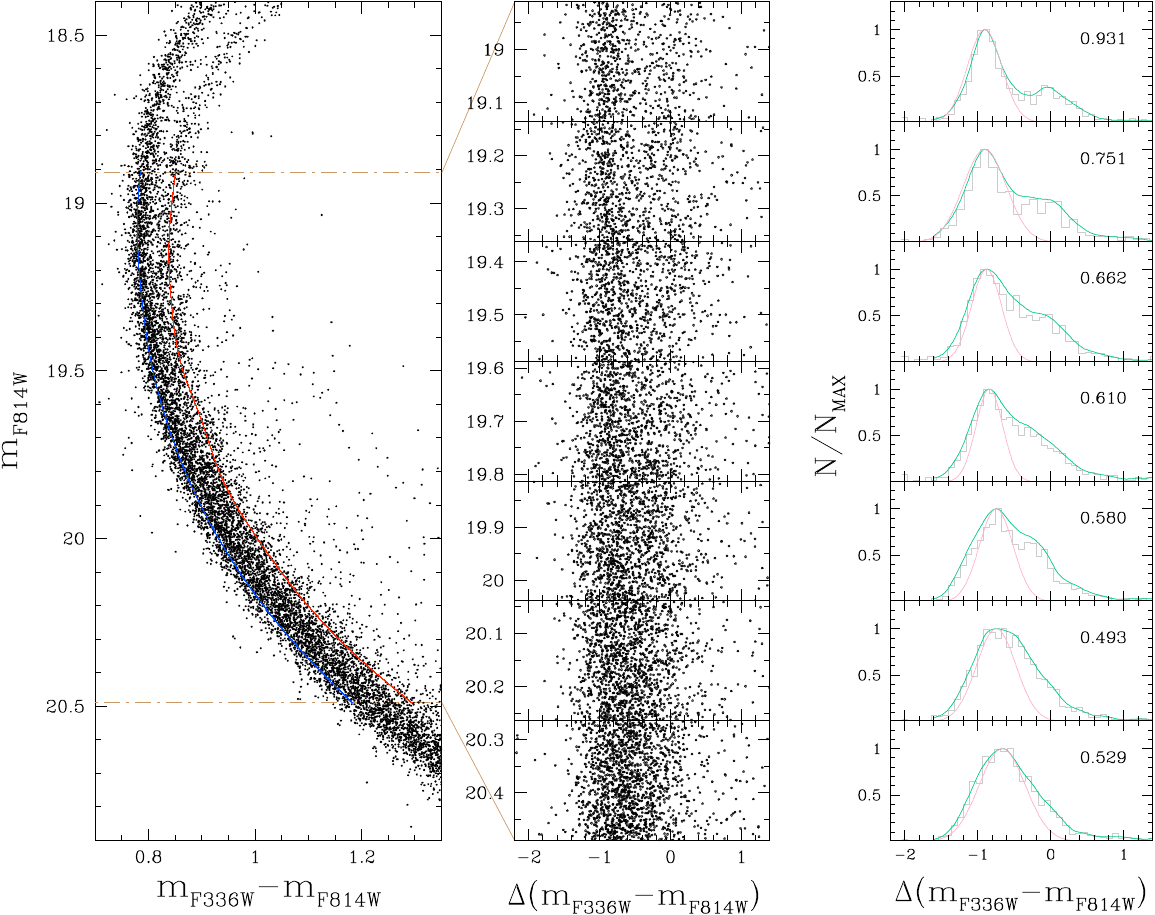}
    \caption{ {\it{Left panel:}} $m_{\rm F814W}$ vs. $m_{\rm F336W}$-$m_{\rm F814W}$ CMD of stars in the {\it HST} catalog outside the innermost 0.7'. Blue and red lines represent the boundaries used to verticalize the color distribution (see text for details), while the brown dot-dashed horizontal lines define the magnitude interval considered in our MS analysis. {\it{Middle panels:}} verticalized $\Delta(m_{\rm F336W}-m_{\rm F814W})$ distribution of MS stars in the $18.9<m_{\rm F814W}<20.5$ interval, divided into 7 magnitude bins. {\it{Right panels:}} $\Delta(m_{\rm F336W}-m_{\rm F814W})$ histogram (in grey) and kernel density (in aqua) distributions of MS stars in each bin defined by the middle panels. The {\bf{pink}} line represents the distribution expected from observational errors.
    }
    \label{fig:mssep}
\end{figure*}

In panels e1) and e2), we plot the $m_{\rm F275W}$-$m_{\rm F336W}$ vs.\,$m_{\rm F336W}$-$m_{\rm F438W}$ two-color diagrams of canonical and anomalous stars, respectively, highlighting in black the SGB stars within the two regions introduced in panel c) and in gray the MS and RGB stars that belong to the same branch (selected by eye).
As proven by \citet{milone2012a}, this two-color diagram is an efficient tool to disentangle stellar populations with different C, N, and O abundances, since the F275W, F336W, and F438W filters encompass the OH, NH, and CH and NH absorption bands, respectively, and it is sensitive to the same chemical variations of the $\Delta_{\rm C F275W, F336W, F438W}$ index in the ChM. 
For that, stars with smaller $\Delta_{\rm C F275W, F336W, F438W}$ (hence larger C and O and smaller N), have smaller $m_{\rm F336W}$-$m_{\rm F438W}$ and larger $m_{\rm F275W}$-$m_{\rm F336W}$ than the stars with larger $\Delta_{\rm C F275W, F336W, F438W}$ (with smaller C and O and larger N). Stellar populations with different C, N, and O abundances form discrete sequences that run parallel on the two-color diagram.
In both panels, each SGB splits in two sequences, which are connected to two separated RGB sequences. By coloring as in Figure~\ref{fig:ellipse} the RGB populations identified in Section~\ref{sec:rgb}, it is clear how the two SGB sequences in panel e1) are the counterparts to the canonical 1G and 2G populations, and that the two anomalous SGBs in panel e2) are linked to the AI and AII RGB stars.
This provides independent confirmation of the quadrimodality observed in Section~\ref{sec:rgb}.

We then apply, to the ground-based $U$ vs. $U$-$I$ CMD, the same procedure to define a sample of canonical and anomalous SGB stars. However, with the available filters, it is not possible to build a diagram able to reveal the four sub-populations, avoiding us to spot the SGB counterpart of 1G, 2G, AI, and AII stars in the outer part of the cluster.

\subsection{Main Sequence}\label{sec:ms}

To investigate the canonical and anomalous MS stars, we consider the $m_{\rm F814W}$ vs. $m_{\rm F336W}$-$m_{\rm F814W}$ CMD and exclude the stars within the innermost 0.7 arcmin to mitigate the effect of crowding on photometry. 
We show a zoom of this CMD on the MS in the left panel of Figure~\ref{fig:mssep}, where the prosecution of the two SGBs is visible in the upper MS, with the bluest and reddest MSs connected to the canonical and anomalous SGBs, respectively.

To further investigate the double MS, we define blue and red boundaries of MS stars between $18.9<m_{\rm F814W}<20.5$ mag and derive their verticalized color $\Delta(m_{\rm F336W}-m_{\rm F814W})$ (see Section~\ref{sec:rgb} for details). The result is plotted in seven different magnitude bins in the middle panels, while in the right panels, we show the histogram (in gray) and the kernel density (in aqua) distribution of $\Delta(m_{\rm F336W}-m_{\rm F814W})$. For each bin, we highlight in pink the distribution of observational errors derived through AS test, arbitrarily centered at the maximum of the kernel distribution, and the Bimodality Coefficient \citep[BC\footnote{$BC = \frac{m_{3}^{2}+1}{m_{4} + 3\frac{(n-1)^{2}}{(n-2)(n-3)}}$, \\
 where $m_{3}$ and $m_{4}$ indicate the skewness of the distribution and its excess of kurtosis, and $n$ is the number of considered points.};][]{sasinc} of the $\Delta(m_{\rm F336W}-m_{\rm F814W})$ distribution of stars. According to the BC criterion, a distribution is considered bimodal if its values exceed the critical threshold $BC_{\rm crit}=0.555$.

Moving from brighter to fainter magnitudes, we notice that: (i) the bimodality becomes less and less clear-cut, as shown by the decrease of the BC, and (ii) the color distribution becomes narrower even if the error increases. These facts agree with two distinct MSs that merge going through fainter magnitudes, with a statistically significant bimodality (i.e., BC>0.555) down to $\sim$20.05 mag.

\section{The radial distribution of multiple stellar populations}\label{sec:fraction}

 \begin{table*}
     \centering
     \begin{tabular}{cc|c|c|c|c|cccc}
                 \hline
                 \hline
                 & & & & & & & & & \\
                 &  & [O/Fe] & [Na/Fe] & [Ba/Fe] & [Fe/H] & &  Fraction      & Fraction      & Fraction \\
                 &  &         &        &         &        & &  (<1.5 arcmin) & (>1.5 arcmin) & (global) \\
                 \hline
                 & & & & & & & & & \\
       CANONICAL & &  0.21 $\pm$ 0.03 &  0.09 $\pm$ 0.04 & 0.46 $\pm$ 0.03 & -1.15 $\pm$ 0.01 & &  0.701 $\pm$ 0.014  &  0.721 $\pm$ 0.031  &  0.705 $\pm$ 0.029  \\
                 & &                    &                    &                   &                    & & (0.706 $\pm$ 0.027) & (0.728 $\pm$ 0.031) & (0.720 $\pm$ 0.030) \\
                 & & & & & & & & & \\
       1G        & &  0.12 $\pm$ 0.03 & -0.12 $\pm$ 0.05 & 0.40 $\pm$ 0.04 & -1.16 $\pm$ 0.01 & & 0.368 $\pm$ 0.018 & 0.437 $\pm$ 0.039 & 0.330 $\pm$ 0.038 \\
       2G        & & -0.05 $\pm$ 0.04 &  0.24 $\pm$ 0.03 & 0.50 $\pm$ 0.03 & -1.15 $\pm$ 0.01 & & 0.632 $\pm$ 0.018 & 0.563 $\pm$ 0.039 & 0.670 $\pm$ 0.038 \\
                 & & & & & & & & & \\
                 \hline
                 & & & & & & & & & \\
       ANOMALOUS & & -0.06 $\pm$ 0.04 &  0.36 $\pm$ 0.05 & 0.74 $\pm$ 0.06 & -1.14 $\pm$ 0.02 & &  0.299 $\pm$ 0.014  &  0.279 $\pm$ 0.031  &  0.295 $\pm$ 0.029  \\
                 & &                    &                    &                   &                    & & (0.294 $\pm$ 0.027) & (0.272 $\pm$ 0.031) & (0.280 $\pm$ 0.030) \\
                 & & & & & & & & & \\
       AI        & &  0.04 $\pm$ 0.04 &  0.16 $\pm$ 0.07 & 0.93 $\pm$ 0.13 & -1.12 $\pm$ 0.02 & & 0.094 $\pm$ 0.027 & 0.229 $\pm$ 0.057 & 0.097 $\pm$ 0.051 \\
       AII       & & -0.10 $\pm$ 0.05 &  0.45 $\pm$ 0.03 & 0.67 $\pm$ 0.06 & -1.15 $\pm$ 0.02 & & 0.906 $\pm$ 0.027 & 0.771 $\pm$ 0.057 & 0.903 $\pm$ 0.051 \\
                 & & & & & & & & & \\
                 \hline
                 \hline
     \end{tabular}
     \caption{Average chemical abundances \citep[from][]{carretta2011}{} of the populations photometrically tagged among RGB stars and their fraction inferred from {\it HST} (within the innermost 1.5 arcmin), ground-based photometry (outside 1.5 arcmin) and over the whole cluster field from its center to the tidal radius.
     Values inside brackets indicate, when present the analogous fraction estimated from the SGB.
     }
     \label{tab:fraction}
 \end{table*}

To investigate the radial distribution of multiple populations in NGC\,1851, we divided the FoV into five (four) circular regions that include the same number of RGB (SGB) stars. We derived the fractions of stars in each population by applying to each region the methods of Sections\,\ref{sec:rgb} and \ref{sec:sgb} for RGB and SGB stars, respectively.

\begin{figure*}
    \centering
    \includegraphics[trim={0 0 0 0},clip,height=8cm]{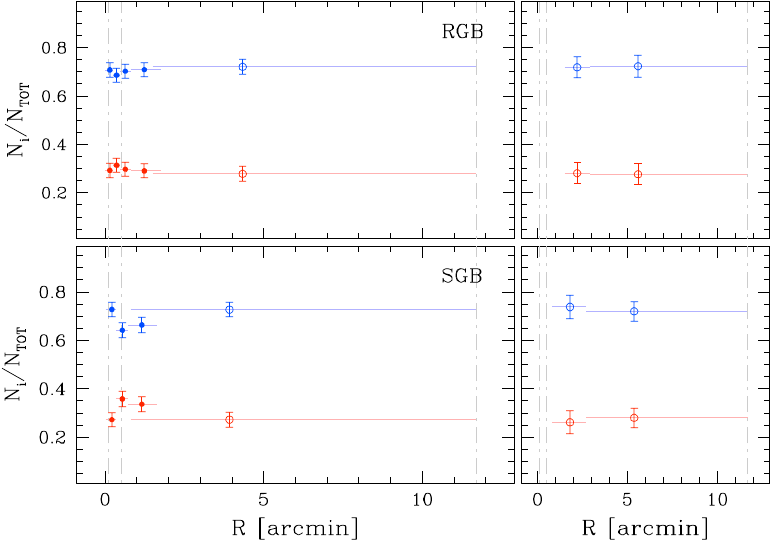}
    \caption{ {\it{Top panels:}} radial trend of the canonical and anomalous RGB star fractions, colored in blue and red respectively, in the {\it HST} and ground-based combined catalog (left panel) and the ground-based catalog only (right panel). Filled and open dots represent measurements obtained from the {\it HST} and ground-based catalog, respectively. {\it{Bottom panels:}} same but for SGB stars. The three vertical dot-dashed lines highlight the core, half-mass, and tidal radius values.
    }
    \label{fig:radial1}
\end{figure*}

As shown in Figure~\ref{fig:radial1}, the fractions of canonical and anomalous stars are constant at the 1-$\sigma$ level over the entire FoV, and such result is obtained from both RGB (top panels) and SGB stars (bottom panels).
We perform a p-value test to infer the probability that the observed behavior is produced by a flat distribution. The derived p-values are 0.92 and 0.29 for RGB and SGB stars, respectively, which strongly support the flat-trend hypothesis (which would be disproved at values $<$0.05).
  
Since the farthest bin covers the whole ground-based radial range, we also consider this catalog alone and divide it into two equal-number bins to explore with higher radial resolution the trend outside $\sim$2 arcmin, as represented in the right panels, finding again no significant radial variation. In each figure, the grey dot-dashed vertical lines represent the core, half-mass, and tidal radius of NGC\,1851\footnote{According to \citet[][2010 edition]{harris1996}, the values of the core, half-mass, and tidal radius are 0.09, 0.51, and 6.52 arcmin, respectively.
To study the cluster halo (see Section~\ref{sec:halo}), we want to be as conservative as possible to select stars that lie outside the tidal radius. For that, we follow the approach by \citet{marino2014} and consider as the tidal radius the largest estimate present in literature, which is from \citet{trager1993} and values 11.7 arcmin.}.

\begin{figure*} [t!]
    \centering
    \includegraphics[trim={0 0 0 0},clip,height=8.2cm]{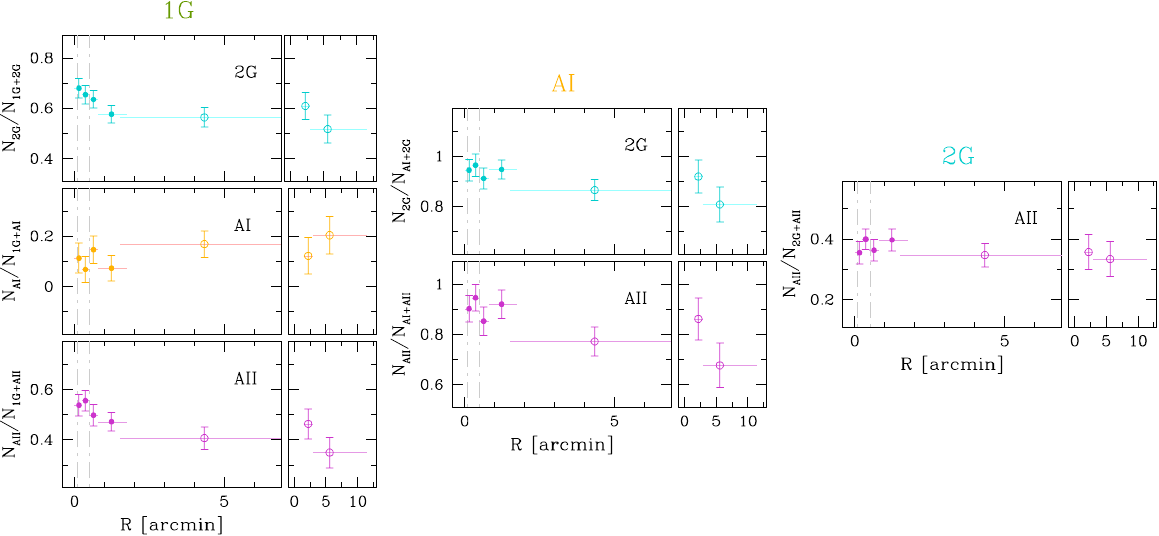}
    \caption{ {\it{Left panels:}} radial distribution of the fraction of the 2G (top), AI (middle), and AII (bottom) populations with respect to the amount of 1G stars.
    {\it{Middle panels:}} same as the right panels but with the 2G and AII populations with respect to AI stars. {\it{Right panel:}} fraction of AII stars with respect to the 2G population.
    }
    \label{fig:rad4}
\end{figure*}

Figure~\ref{fig:rad4} explores the radial distributions of the four populations. We considered different pairs of populations and derived their fractions in equal-number bins. 
Moreover, we further divided the ground-based field into two circular regions with the same number of stars. Results are shown in Figure~\ref{fig:rad4}. The left column represents the radial behavior of the 2G, AI, and AII star fractions to the 1G population, revealing that 2G and AII are more centrally concentrated than 1G stars, while no variation appears between 1G and AI stars. The central column compares the AI population, to 2G and AII stars. Their ratios decrease when moving through larger radii, suggesting that AI stars, such as the 1G, are more diffused than the 2G and AII stars. Finally, in the right column, we compare the 2G with the AII population, detecting no radial difference.

Finally, we derive the global fraction of the different populations spotted in NGC\,1851. To do that we convolve, from the center to the tidal radius, the radial trends illustrated in Figures~\ref{fig:radial1},~\ref{fig:rad4} by the best-fit King profile \citep{king1962} derived by \citet[][2010 edition]{harris1996} to account for the radial density distribution of the cluster stars.
Our resulting fractions, derived from RGB stars, are listed in Table~\ref{tab:fraction}.
To estimate the uncertainties, we simulate 10,000 radial distributions by scattering the observed radial trends by their errors. Then, we repeat the procedure to infer the global ratios for each sample and considered as our uncertainties the 68-th percentiles of the distribution of the global fraction obtained from all the simulations is our uncertainty.

\section{spatial distribution} \label{sec:spatial}

To investigate the 2D spatial distribution of multiple populations, we applied to the canonical and anomalous RGB and SGB stars the method by \citet[][see their Section 3]{cordoni2020}, which is based on a 2D kernel smoothing of the coordinate distribution ($\Delta$RA and $\Delta$DEC).
 
\begin{figure*}
    \centering
    \includegraphics[trim={0 -3 0 0},clip,height=13cm]{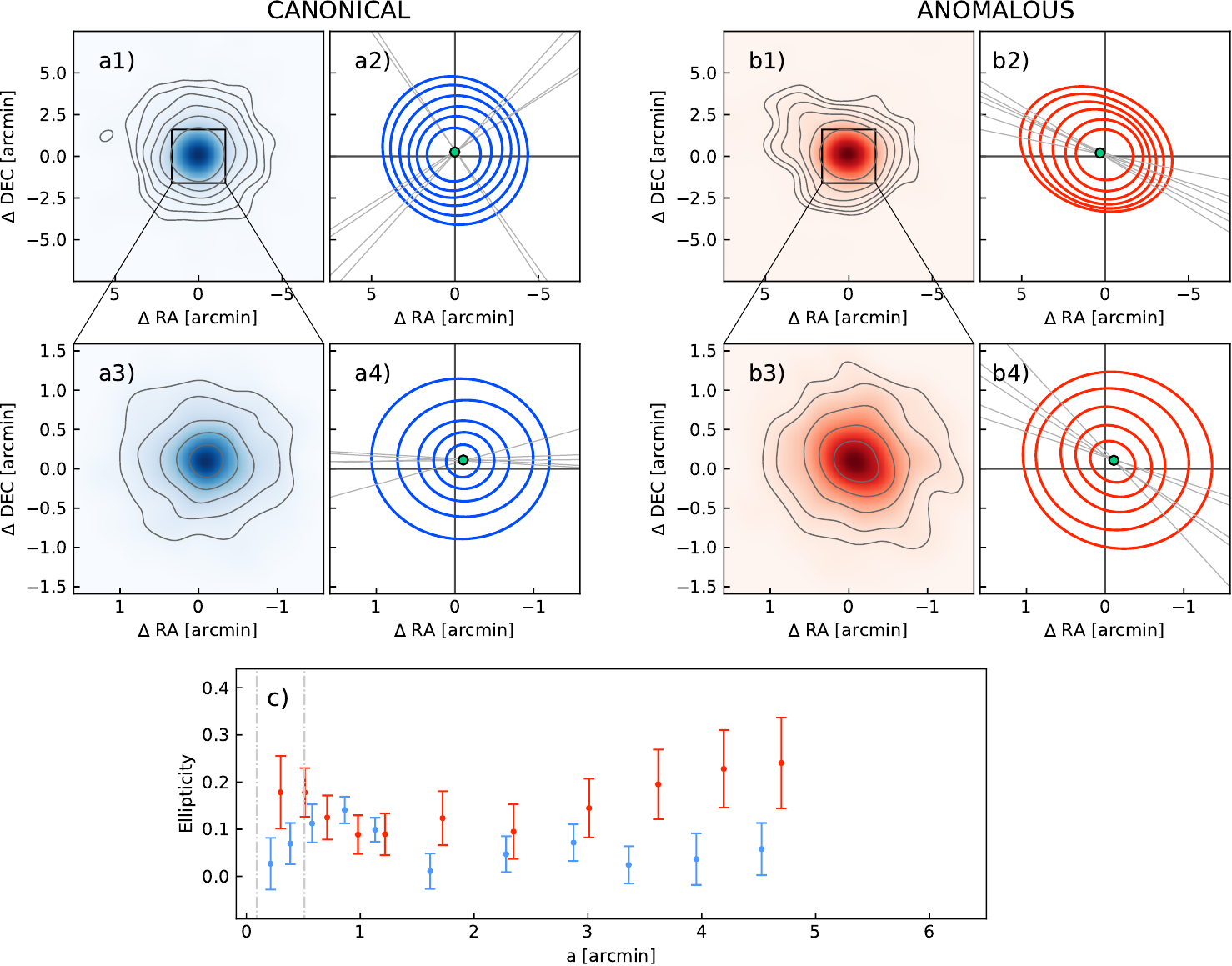}
    \caption{ {\it{Panels a1) and b1):}} spatial distribution of canonical and anomalous stars in the ground FoV, represented in blue and red color scale, respectively. Dark grey lines are the isodensity contours. {\it{Panels a2) and b2):}} best-fit ellipses of canonical (in blue) and anomalous (in red) isodensity lines. Grey straight lines represent the major-axis direction of each ellipse, while aqua dots display the averaged ellipse centers. {\it{Panels a3), b3), a4), and b4):}} same as panels a1), b1), a2), and b2) but for stars in the {\it HST} FoV (within the black boxes in panels a1) and a2)).
    {\it{Panel c):}} Ellipticity of canonical and anomalous stars with respect to the major axis of their isodensity contours best-fit ellipses. The two dash-dotted lines represent the core and the half-mass radius. 
    }
    \label{fig:2dsp}
\end{figure*}

The resulting smoothed 2D distribution of canonical and anomalous stars in the ground-based catalog are portrayed in panels a1) and b1) of Figure~\ref{fig:2dsp}, respectively. We then compute their isodensity lines and fit them with ellipses by means of least-squares by using the \citet{halir1998} algorithm. The best-fitting ellipses are displayed in panels a2) and b2), where we highlight the major axis of each of them (grey lines) and the average center (aqua bullet).
Panels a3) and a4) represent the 2D kernel-smoothed distribution in the innermost $\sim$1.5 arcmin, obtained with {\it HST} data and the corresponding best-fitting ellipses.

This figure highlights some differences between the two populations.
Canonical stars exhibit a nearly circular distribution over the entire FoV and their position angle are poorly constrained. Conversely, the anomalous population shows a circular distribution in the innermost areas only, whereas for radial distances larger than $\sim 3.5$ arcmin the anomalous stars exhibit higher ellipticity values than  the canonical stars\footnote{The ellipticity is defined as $1-\frac{b}{a}$, where $a$ and $b$ are the major and minor axis, respectively.}. All the best-fitting ellipses that describe the anomalous stars share similar orientations with an average position angle of $\sim$30$^{o}$.
These results are illustrated in panel c), in which we plot the ellipticity of each best-fitting ellipse against its major axis, showing that outside $\sim$3.5' the anomalous population is more elliptical at a 1-$\sigma$ level than the canonical one. 
The uncertainty associated with the ellipticity is derived by simulating 1,000 sample of stars with the same number of stars and spatial distributions as canonical and anomalous stars. For each simulation, we measured the ellipticities with the same method used for the real stars. 
The error associated with each measurement is then calculated as the 68$^{\rm th}$ percentile of each simulated ellipticity distribution.
Finally, we notice an overdensity of anomalous stars in the north-eastern quadrant, that forms the elongation in the contour plot in panel b1) around $(\Delta RA, \Delta DEC) \sim (3.5, 3.0)$ arcmin. Indeed, the fraction of ground-based-field anomalous stars in this quadrant is significantly higher than average, being 0.38$\pm$0.05 against the overall 0.28$\pm$0.03.

We repeat the analysis by considering the 1G and 2G canonical subpopulations in the RGB (the separation between these two populations among SGB stars is not clear enough for a reliable quantitative analysis). 
The low number of stars leads to poor statistics, hence large uncertainties in our ellipticity measurements. With that in mind, we still point out that no significant difference between their average ellipticity emerged between the two spatial distributions, obtaining 0.128$\pm$0.069 and 0.086$\pm$0.62 for the 1G and 2G stellar populations, respectively.

\subsection{Stars outside the tidal radius} \label{sec:halo}

An intriguing feature of NGC\,1851 is the presence of a halo of extratidal stars that surrounds the cluster up to $\sim$500 pc, as discovered by  ~\citet[][]{olszewski2009} and further explored over larger scales by \citet{carballo2018, kuzma2018, ibata2021}.

\begin{figure*}
    \centering
    \includegraphics[clip,height=14.5cm]{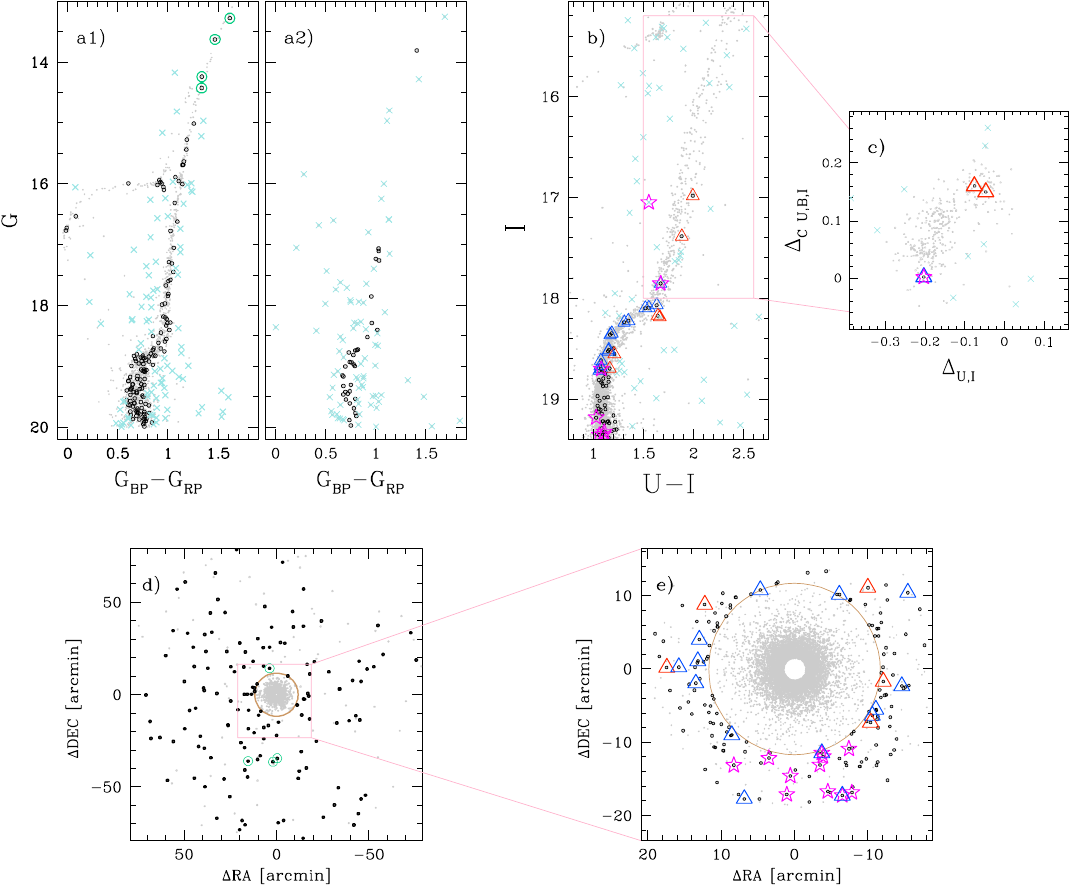}
    \caption{ {\it{Panel a1) and a2):}} Gaia $G$ vs. $G_{\rm BP}$-$G_{\rm RP}$ CMD of stars within 80 arcmin from the cluster center and from the FoV dominated by field stars, respectively. Stars that pass the photometric diagnostics and the proper motion selection are marked with gray points. Black points and azure crosses represent the extratidal stars that are consistent with belonging to NGC\,1851 and to the field according to the CMD selection, respectively. Stars with radial velocity measurements consistent with the cluster motion are encircled in aqua (see text for details).
    {\it{Panel b):}} $I$ vs. $U$-$I$ CMD from ground-based photometry. Grey and black points and azure crosses have the same meaning than in panel a1) and a2). Halo canonical and anomalous stars are displayed with blue and red triangles, respectively, while magenta starred-symbols display the stars in common with the work by \citet{marino2014}.
    {\it{Panel c):}} $\Delta_{\rm C U,B,I}$ vs. $\Delta_{\rm U,I}$ ChM for RGB stars within the pink box in panel b).
    {\it{Panel d):}} $\Delta$DEC vs. $\Delta$RA position of stars in the Gaia FoV, color-coded as in panel a1) and a2).
    {\it{Panel e):}} zoom of the Gaia $\Delta$DEC vs. $\Delta$RA diagram within the pink rectangle representing the position of stars in the ground-based FoV. The brown circle in panels d) and e) indicates the tidal radius.
    }
    \label{fig:halo} 
\end{figure*}

The recent Gaia DR3 data allows us to reach the outermost areas of the cluster, exploring its stellar halo. We apply to the Gaia catalog the following criteria to identify halo stars: (i) we consider only sources with $G<$20 to exclude the ones with low signal-to-noise ratio. (ii) we use the {\tt{astrometric\_gof}}, the renormalized unit weight error ({\tt{RUWE}}), and the parallax diagnostics provided in the Gaia catalog to select only the sources with high-quality photometry. (iii), we analyze the Gaia proper motions and select the stars within a radius of 0.9 mas yr$^{\rm -1}$ centered on the average proper motion. (iv) we select by eye in the $G$ vs. $G_{\rm BP}$-$G_{\rm RP}$ CMD the stars that lie on the MS-SGB-RGB-HB evolutionary sequence, therefore that they are reasonable cluster members. 
We show, in panel a1) of Figure~\ref{fig:halo}, the $G$ vs. $G_{\rm BP}$-$G_{\rm RP}$ CMD of NGC\,1851. We highlight in grey all the stars that fulfill (i), (ii), and (iii), while stars in the halo (i.e., outside the tidal radius) that pass also the CMD selection criterion are represented with black points. Azure crosses indicate the stars outside the tidal radius that, according to our fourth criterion, are excluded from belonging to the cluster.

We consider a circular region of the sky extended up to 80 arcmin ($\sim$260 pc) from the cluster center, and after applying these strict selection criteria we still detect NGC\,1851 stars all over this FoV, thus confirming of the presence of stars outside the tidal radius. Furthermore, for some of the black stars, there are also available radial velocity measurements in the Gaia DR3, which can serve as a further diagnostic for cluster membership. Based on the spectroscopy results in literature \citep[e.g.,][]{sollima2012, marino2014}, we consider as cluster members the stars with a radial velocity between 300 and 350 km s$^{\rm -1}$. The ones that, beyond respecting the aforementioned selection criteria, also fulfill the radial velocity criteria, are encircled in aqua, and can be observed up to a radius of $\sim$38'.

To investigate the contamination of field stars with proper motions and CMD position similar to NGC\,1851 stars in our sample of halo stars, we consider an annulus with the same area as the halo field but located further from the cluster center, between 140 and 160 arcmin, where we expect a negligible presence of cluster stars. The CMD of this field is represented in panel a2).
From the tidal radius up to 80 arcmin, we find 140 halo stars and 1,256 field stars. In the outer annulus, the number of field stars is comparable (1,401), while only 35 stars share the same colors, magnitude and proper motions as cluster members. These results prove that our sample of halo stars is not consistent with being made by field stars only (as in the outer annulus). Specifically, by assuming an uniform distribution in the sky, we expect that the contamination from field stars is about 25\% (~35/140).

We use ground-based photometry to identify extratidal canonical and anomalous stars and explore their distribution along the FoV. The FoV with available $U$ and $I$ photometry covers a part of the NGC\,1851 halo only. Specifically, the catalog includes stars within $\sim$20 arcmin at east and south, and within 13 and 18 arcmin at north and west, respectively, that pass the criteria of selection described in Section~\ref{sec:data}.

A similar investigation was performed by \citet[][]{marino2014}, who inferred s-process elements abundances through spectroscopy in the southern area of the halo, finding that 15 of these stars have radial velocity and metallicity consistent with the cluster and that they all share the s-process elements abundances with the canonical stars. We show in the panel b) of Figure~\ref{fig:halo} the $I$ vs. $U-I$ CMD, where black points and azure crosses represent the extratidal stars that we included and excluded in our sample of cluster stars, respectively, based on their CMD position.
Magenta starred-symbols highlight the halo stars of the Marino and collaborators sample that are present in our catalog and pass the photometric quality criteria. Canonical and anomalous halo SGB and RGB stars are indicated with blue and red open triangles, respectively. While the former are identified by their position on the CMD, the latter were classified through the $\Delta_{\rm C U,B,I}$ vs. $\Delta_{\rm U,I}$ ChM (in panel c)), derived as in Section~\ref{sec:rgb}, extended down to I$=$18 mag. Here, we identify one probable canonical RGB star, consistent with belonging to the 1G population, and two anomalous AII stars. We consider only these three stars as reliable canonical and anomalous RGB stars. 
We calculate the coordinates, relative to the cluster center, $\Delta$DEC and $\Delta$RA, and show in the panel d), the $\Delta$DEC vs. $\Delta$RA diagram of the halo population that we identified from Gaia data. Finally,  panel e) illustrates a zoom of the region covered by well-measured stars from the \citet{stetson2019} catalog. In agreement with the result from \citet{marino2014}, we do not detect anomalous stars in the southern part of the halo below $\Delta DEC \sim$ $-$10 arcmin  (with two of these stars being tagged both by spectroscopy and photometry), while along other directions we identified five probable anomalous stars.

This result, even though based on a limited number of stars, suggests an uneven distribution of anomalous stars in the halo, with a lack of them in the south and southeast directions. Noticeably, this is qualitatively consistent with the findings shown in Figure~\ref{fig:2dsp}, where the anomalous population is less extended along these directions.

\section{Summary and conclusions}\label{sec:final}

We used multi-wavelength photometry from {\it HST}, Gaia DR3, and ground-based facilities to disentangle and characterize the stellar populations of the Type\,II GC NGC\,1851.
The multiple populations are analyzed over a wide area that ranges from the cluster center to the outskirts. Our main results are summarized in the following:

\begin{itemize}

    \item Both {\it HST} and ground-based photometry reveal that the distribution of stars along the ChM  of both canonical and anomalous RGB stars is bimodal.
    The canonical population comprises the s-poor stars  while the anomalous population hosts the s-rich stars discovered by \citet[][]{yong2008}.
    The canonical and anomalous stars can be followed continuously along the RGB, SGB, and upper MS, where they define two distinct sequences that merge around one F814W magnitude below the turnoff. 

    \item
     Based on the ChMs, we identified the stellar populations within the canonical and anomalous populations.
     The canonical population hosts the distinct groups of 1G and 2G stars, typically observed in Type I GCs. These two populations have different abundances of helium, carbon, nitrogen, and oxygen.
     Both 1G and 2G stars are not chemically homogeneous. Similarly, the anomalous RGB hosts two main populations, AI and AII, with different light-elements abundances. 
    
    \item To constrain the overall CNO abundance of canonical and anomalous stars we compared their observed colors with the colors derived from synthetic spectra with different contents of carbon, nitrogen, and oxygen. We found that canonical and anomalous stars share similar average abundances of carbon, while the anomalous stars are enhanced in [N/Fe] by $\sim$1.0 dex and slightly depleted in oxygen by $\sim$0.1 dex. Hence, the anomalous stars have enhanced CNO content by 0.35$\pm$0.10 dex with respect to the canonical population.
    Our results, which are based on multi-band photometry, confirm the findings by \citet{yong2015}, who obtained similar conclusion by using high-resolution spectra.
    
    \item 
    We investigated the radial distribution of the distinct population fractions up to the tidal radius. 
    We find that the canonical and anomalous stars share the same radial distribution. We instead found that 2G stars are more centrally concentrated than the 1G, and AII stars are more centrally concentrated than AI stars.
    We did not detect significant differences between the 1G and AI, and the 2G and AII.
    
    \item
    We then exploited the radial trend of the different population fractions to measure the global fraction of the different populations reported in Table~\ref{tab:fraction}. The global fractions of the four disentangled populations with respect to the total number of canonical and anomalous stars are $f^{\rm G}_{\rm 1G} =$ 0.229 $\pm$ 0.030, $f^{\rm G}_{\rm 2G} =$ 0.474 $\pm$ 0.030, $f^{\rm G}_{\rm AI} =$ 0.027 $\pm$ 0.030, and $f^{\rm G}_{\rm AII} =$ 0.270 $\pm$ 0.030.

    \item Canonical and anomalous stars differ in the 2D spatial distributions.
     The isodense contours of canonical stars have nearly circular shapes (ellipticity of $\sim$0.1) in the entire FoV.
     The contours of anomalous stars deviate from a circular-like shape outside the innermost three arcmin increasing in ellipticity up to $\sim$0.3 and having its best-fit ellipse oriented along the north-east/south-west direction. Moreover, there is a hint for an overdensity of anomalous stars in the northeast direction, where their fraction increases by $\sim$10\% with respect to the average, which, as shown in Section~\ref{sec:fraction}, is significant at a 2$\sigma$ level.

    By combining the analysis of the radial and spatial 2D distribution of canonical and anomalous stars, we found that their overall fractions do not vary within the tidal radius, in agreement with the findings by \citet{milone2009}, but the uneven distribution of the anomalous population introduces local gradients. In particular, their drop in the south/southeast outer field of the cluster is qualitatively consistent with the results by \citet{zoccali2009}, who detected a gradient by studying a similar field.

    \item We identify NGC\,1851 stars outside the tidal radius, thus confirming previous results \citep[][]{olszewski2009, sollima2012, marino2014, kuzma2018}. By using Gaia DR3 data, we detect a stellar halo up to 80 arcmin from the cluster center.
    We identified 14 canonical and five anomalous probable cluster members outside the tidal radius (radial distances between $\sim$12 and $\sim$20 arcmin) thanks to the available ground-based photometry.  
    The tagging of canonical and anomalous stars outside the tidal radius corroborates the idea that anomalous stars nearly disappear along the south/southeast direction \citep{marino2014}, but are still visible in other directions.
    Since the available observations allow us to separate these two populations up to about 20 arcmin, and the halo is extended up to (at least) 80 arcmin, it is clear how extending this analysis to higher radii is mandatory to shed light on this phenomenon.
    
\end{itemize}

We conclude by providing some considerations, although strictly qualitative, about the formation of anomalous stars in Type II GCs. Two main ideas are particularly appealing based on our observational constraints.

The first one predicts that Type II clusters result from a merging between two (or more) initially separated Type I GCs \citep[][]{carretta2010, bekki2016}. According to this idea, they form within the same dwarf galaxy, develop their own 1G-2G patterns, and then spiral in the nuclear region of the host galaxy, merging in one. Finally, the galaxy is accreted by the Milky Way, which strips its stars leaving only the naked nuclei, i.e., the Type II GC.
Here, the differences between the canonical and anomalous stars chemistry would arise as a result of the dwarf galaxy chemical evolution. Indeed, the cluster in which anomalous stars were born would have formed later, thus when the star-forming gas in the host galaxy had a different chemical composition. \citet{bekki2016} performed simulations to show that the iron and s-process differences observed in M22 could be achieved within a few hundred of Myr, hence before the dwarf disruption.
This idea naturally accounts for the presence of two anomalous populations, AI and AII, which would be the first and second generation of an initially-separated GC. Moreover, the 1G-2G and AI-AII patterns also share similar relative chemical differences and radial distribution, which would indicate them being produced by the same mechanisms.
On the other hand, the large C$+$N$+$O difference observed between canonical and anomalous stars in NGC\,1851 is not straightforward in this scenario, which would require excessively long timescales to produce it \citet[][see their Section 4.2]{bekki2016}.
Finally, we notice how AI stars would have a rather extreme chemical composition for being first-generation stars, having intermediate [Na/Fe] and [O/Fe] between 1G and 2G stars, and it is not clear if the chemical evolution of a host dwarf galaxy could account, in the required timescales, to such chemical differences.

The second scenario, proposed by \citet[][]{dantone2016} and \citet[][]{dercole2016} is an extension of the AGB scenario \citep[e.g.,][]{dercole2008}. Here, Type II GCs experienced a prolonged star formation with respect to Type I GCs, allowing subsequent stellar generations to form.
After the formation of 2G stars, the explosions of delayed SN II in binaries destroy the cooling flow and halts the stellar formation, but because their frequency is not as high as in the single SN II epoch, they are not strong enough to push the intra-cluster medium (formed by pristine material and AGB ejecta) out of the cluster proximity. Type II GCs, differently from Type I, can re-accrete this gas several Myr later, when the delayed SN II events become rare. This could be possible by assuming that these clusters were particularly massive at this epoch or if they are nuclei of disrupted dwarf galaxies.
The re-accreted material would be contaminated by SN II ejecta, thus enriched in iron. In this time span, $\sim$3.5-4 $M_{\rm \odot}$ AGB stars are polluting the winds injecting material strongly affected by the third dredge-up, hence enriching the surroundings in total CNO and s-process elements. 
In this mixed medium, anomalous stars would form and would be enriched in total CNO, s-process elements, and/or [Fe/H] depending on the influence of the different polluters within a given GC (like the number of delayed SN II events). 
If the mixing between different ejecta is inhomogeneous, it is possible to develop a Na-O anticorrelation among anomalous stars \citep[][see their Section 4.2]{dercole2016}, thus producing the observed AI and AII populations.

Both scenarios agree on the possibility that Type II GCs may be remnants of a larger structure, like a dwarf galaxy. The presence of a halo of stars more extended than the tidal radius of NGC\,1851 could be a sign that this cluster was originally a larger structure. An extensive study of the halo, aimed at identifying the populations that compose it, will provide additional constraints on the origin of NGC\,1851 and, possibly, of other Type II GCs.

Our results provide new constraints and challenges to the Type II GC formation scenarios. To unveil the origin of these structures further works based on investigating anomalous stars in a wider sample of clusters, combining photometry, spectroscopy, and theoretical modeling, are mandatory.

\section*{Acknowledgements}

We thank the anonymous referee for the valuable comments.
This work has received funding from the European Research Council (ERC) under the European Union’s Horizon 2020 research innovation programme (Grant Agreement ERC-StG 2016, No 716082 ’GALFOR’, PI: Milone, http://progetti.dfa.unipd.it/GALFOR) and from the European Union’s Horizon 2020 research and innovation programme under the Marie Sklodowska-Curie Grant Agreement No. 101034319 and from the European Union – NextGenerationEU, beneficiary: Ziliotto. SJ acknowledges support from the NRF of Korea (2022R1A2C3002992, 2022R1A6A1A03053472).  APM, MT, and ED acknowledge support from MIUR through the FARE project R164RM93XW SEMPLICE (PI: Milone). APM and ED have been supported by MIUR under PRIN program 2017Z2HSMF (PI: Bedin). FD and PV acknowledge the support received from the PRIN INAF 2019 grant ObFu 1.05.01.85.14 (“Building up the halo: chemo-dynamical tagging in the age of large surveys”, PI. S. Lucatello) and the INAF-GTO-GRANTS 2022 (“Understanding the formation of globular clusters with their multiple stellar generations”, PI. A.\,F.\,Marino).
ZO acknowledges this research was supported by an Australian Government Research Training Program (RTP) Scholarship. AK and ZO were supported by the Australian Research Council Centre of Excellence for All Sky Astrophysics in 3 Dimensions (ASTRO 3D), through project number CE170100013.

\section*{Data Availability}

The data underlying this article will be shared upon reasonable request to the corresponding author.



\bibliographystyle{mnras}
\bibliography{main} 








\bsp	
\label{lastpage}
\end{document}